\documentclass{emulateapj}

\newcommand{\myemail}{quanz@astro.phys.ethz.ch}

\shorttitle{Very low-mass brown dwarfs and planetary-mass objects in Taurus}
\shortauthors{Quanz et al.}

\begin{document}


\title{Search for very low-mass brown dwarfs and free-floating planetary-mass 
objects in Taurus}


\author{Sascha P. Quanz}
\affil{Institute for Astronomy, ETH Zurich, Wolfgang-Pauli-Strasse 27, 8093 Zurich, Switzerland}
\affil{Max Planck Institute for Astronomy, K\"onigstuhl 17, Heidelberg,
    Germany}
\email{\myemail}
\author{Bertrand Goldman, Thomas Henning, Wolfgang Brandner}
\affil{Max Planck Institute for Astronomy, K\"onigstuhl 17, Heidelberg,
    Germany}
\author{Adam Burrows}
\affil{Department of Astronomy and Steward Observatory, University of Arizona, Tucson, AZ 85721}
\affil{Department of Astrophysical Sciences, Peyton Hall, Princeton University, Princeton, NJ 08544}
\and
\author{Lorne W. Hofstetter}
\affil{Department of Physics, Princeton University, Princeton, NJ 08544}
\altaffiltext{1}{Based on observations made at the Calar Alto Observatory. Based on observations made 
with ESO Telescopes at the Paranal Observatories under program ID 278.C-5043A. 
This work is based in part on observations made with the {\sc Spitzer Space Telescope}, which is operated by the Jet Propulsion Laboratory, California Institute of Technology under a contract with NASA.}


\begin{abstract}
The number of low-mass brown dwarfs and even free floating planetary mass objects in young nearby star-forming regions and associations is continuously increasing, offering the possibility to study the low-mass end of the IMF in greater detail. In this paper, we present six new candidates for (very) low-mass objects in the Taurus star-forming region one of which was recently discovered in parallel by \citet{luhman2009caha6}. The underlying data we use is part of a new database from a deep near-infrared survey at the Calar Alto observatory. The survey is more than four magnitudes deeper than the 2MASS survey and covers currently $\sim$1.5 deg$^2$. Complementary optical photometry from SDSS were available for roughly 1.0 deg$^2$. After selection of the candidates using different color indices, additional photometry from Spitzer/IRAC was included in the analysis. In greater detail we focus on two very faint objects for which we obtained $J$-band spectra. Based on comparison with reference spectra we derive a spectral type of L2$\pm$0.5 for one object, making it the object with the latest spectral type in Taurus known today. From models we find the effective temperature to be 2080$\pm$140 K and the mass 5-15 Jupiter masses. For the second source the $J$-band spectrum does not provide a definite proof of the young, low-mass nature of the object as the expected steep water vapor absorption at $1.33\,\mu$m is not present in the data. We discuss the probability that this object might be a background giant or carbon star. If it were a young Taurus member, however, a comparison to theoretical models suggests that it lies close to or even below the deuterium burning limit ($<$13 M$_{Jup}$) as well. A first proper motion analysis for both objects shows that they are good candidates for being Taurus members. 
\end{abstract}





\section{Introduction}
Taurus is certainly one of the best studied low-mass star-forming (SF) regions in the northern hemisphere. It lies
at a distance of $\sim$\,140 pc \citep{kenyon1994} and is very young \citep[median age $\sim$1 Myr,][]{briceno2003,luhman2003b}. The initial mass function (IMF) 
in Taurus has been studied intensively in the last years using large surveys mostly in the optical. 
It first seemed as if at the low-mass end the number of brown dwarfs (BDs) was too small compared to the expectations for a typical IMF \citep{briceno2002}. While other studies found indications that this might not be the case \citep{luhman2006spitzer,guieu2006} a recent analysis finds that a surplus of late K and early M type stars seems still to be present \citep{luhman2009xray}. Apart from one exception the least massive objects in Taurus known today have spectral types of M9-M9.5 with corresponding effective temperatures of $\sim$\,2400-2500 K \citep[][and references therein]{luhman2009caha6,luhman2006distribution,guieu2006}. Table~\ref{bd_masses1} shows that theoretical models predict these objects to have masses above the deuterium burning limit of $\sim$0.013 M$_{\sun}$$\approx$13 M$_{Jup}$ \citep{burrows1997} which makes them very low-mass brown dwarfs. Only very recently \citet{luhman2009xray} reported the detection of the first L0 dwarf in Taurus where the spectral type was derived from an I-band spectrum. However, this object appears to be underluminous for its assumed young age and we will discuss this object and its properties in more detail in section~\ref{discussion}. In theory, young objects (1-3 Myr) with L and T spectral types and corresponding effective temperatures below $\sim$2300 K could potentially be of planetary mass (Table~\ref{bd_masses2}).   

For other young star-forming regions the detection of L- and T-type objects has already been reported and their numbers are rather small but continuously increasing. In the 3-5 Myr old cluster $\sigma$-Orionis one object, S Ori 70, has an observed spectral type of T5.5$\pm$1 \citep{zapateroosorio2002}, one object is of spectral type L4 and three objects have spectroscopically confirmed spectral types of L0-L1.5  \citep{zapateroosorio2000}. If these objects are members of the young cluster then they can be considered as very good candidates for isolated planetary mass objects (i.e., with masses M$<$13 M$_{Jup}$). However, \citet{mcgovern2004} found for one potential candidate in $\sigma$-Orionis (S Ori 47) spectroscopic evidence that the object is rather an object from the field than from the young cluster. Recently, \citet{bihain2009} report the detection of three additional objects in $\sigma$-Orionis (S Ori 72-74) with theoretical masses of a few Jupiter masses (with L-T spectral types), but spectroscopic confirmation is still pending. In addition to the objects in $\sigma$-Orionis \citet{lodieu2008} and \citet{luhman2008chameleon} report the finding of L-type objects in the Upper Sco Association and the Chameleon I region, respectively (median ages $\sim$3-5 Myr). \citet{weights2009} report the spectroscopic confirmation of 38 very low mass objects below the hydrogen burning limit in the Orion Nebular cluster (median age $\sim$1 Myr). 10 of their objects could have masses below the deuterium burning limit and be thus planetary mass objects. \citet{burgess2009} recently published the detection of a T-dwarf candidate in the 3 Myr-old star-forming region IC 348 based on methane-band imagery. The found a preliminary spectral type of T6 for the object and concluded that the detection of this source is consistent with the extrapolation of current lognormal IMF estimates down to the planetary mass domain given the survey area and completeness of their sample. 
Furthermore, there are three optically classified L dwarf candidates in the field showing some spectral signatures of youth \citep{kirkpatrick2001,kirkpatrick2006,luhmanwilson2006}. Four additional objects in young  star-forming regions (1-3 Myr) are at the borderline between M and L-type objects but are subject to controversial discussions in the literature: \citet{jayawardhana2006a,jayawardhana2006b} derived for four objects in Chameleon, Lupus and Ophiuchus spectral types of L0 and planetary masses while \citet{luhman2007astroph} and \citet{allers2007} derived at least for three of the objects significantly earlier spectral types (M7.25-M8.75) and higher masses. In the and the TW Hya association (5-10 Myr) only late M-type and thus more massive objects were found \citep{liu2003,gizis2002}. 


In this paper we present new candidates for low-mass BD in Taurus. At least two objects are good candidates for being among the least massive objects known today in this SF region possibly lying below or close to the edge of isolated planetary mass objects. For one object we derive a spectral type of L2$\pm$0.5 and a most likely mass between 5 and 15 Jupiter masses.

\section{Observations and data reduction}\label{observations}
\subsection{{\sc Omega2000} data} 
In a new deep near-infrared (NIR) survey of Taurus we focused on regions of comparably high extinction where previous optical surveys might have missed potentially young, very low-mass objects.
In the first part of our ongoing survey we observed the vicinity of 55 dense molecular cores \citep{onishi2002} between October 
2004 and January 2006 with the wide-field NIR 
camera {\sc Omega2000} on the 3.5 meter telescope in Calar Alto (Spain)\footnote{http://w3.caha.es/CAHA/Instruments/O2000/index2.html}. 
The camera is equipped with a 2048$\times$2048 pixel HAWAII-2 detector,
has a pixel scale of 0.45$''$/pixel and an effective field-of-view of $\sim$15$'$. The 55 dense cores were covered with 24 telescope pointings. One additional field harboring a proto-brown dwarf \citep{apai2005} was included in the survey. Up to now the NIR observations cover an area of $\sim$1.5 deg$^2$ on the sky (Table~\ref{o2kobservations}). Each field was observed in the three NIR filters $J$, $H$, and $K_s$ with central wavelengths at 1.21, 1.65, and 2.15\,$\mu$m, respectively. For two fields the $K_s$-band data had to be disregarded in the data analysis as cloudy skies disturbed the observations significantly. Following a pre-defined dither pattern 30 frames were taken on each field and filter. Each frame had an on-source integration time of one minute, so that after co-addition the total integration time amounted to 30 minutes per field and filter. The limiting magnitudes were $\sim$\,22.0 mag in $J$, $\sim$\,20.5 mag in $H$ and $\sim$\,19.5 mag in $K_s$. The depth of the survey allows us to detect young objects (1-5 Myr) down to a few Jupiter masses in all three filters if they were present.

The data reduction was carried out with a dedicated pipeline developed at the MPIA. The pipeline runs within the
\texttt{MIDAS} environment and does flat-fielding, bad pixel and cosmic ray removal, sky subtraction and finally weighting and co-addition of the
individual images to a combined final image for each field and filter. If one frame suffers substantially from bad quality, e.g., due to bad observing conditions in terms of seeing or transparency, this frame is rejected and not included in the final image. A more detailed overview concerning the
pipeline is provided in the {\sc Omega2000} Handbook\footnote{http://www.caha.es/CAHA/Instruments/O2000/index.html}.

Astrometry was applied to each field by comparing the positions of detected 2MASS point sources to those listed in the
2MASS Point Source Catalog (PSC) \citep{cutri2003}. On average the precision of the astrometry is better than 0.25$''$ with respect to 
the known 2MASS positions.

After visual inspection of the resulting final frames, photometry was carried out with the \texttt{daophot} package within the 
\texttt{IRAF}\footnote{http://iraf.noao.edu/} environment. As the point spread function (PSF) does not
vary significantly over the chip, PSF-photometry was carried out instead of simple aperture photometry, in order to ensure reasonable flux measurements of close binaries and in more crowded regions. A reference PSF 
was created for each filter and field individually and then 
applied to all sources detected with a 3-$\sigma$ confidence level. 
The resulting tables containing the fluxes for each field and filter were then cross-matched 
in order to identify sources that were detected in all three filters. For the cross-matching the coordinates of the 
measured flux peaks were required to match within 2 pixels. 

Finally, the photometric calibration was achieved by comparing the measured 
instrumental magnitudes to those listed in the 2MASS PSC. Here, only objects with the best "quality flag" and "read flag" in the 2MASS PSC were considered serving as photometric reference sources. Typically between 50 and 100 objects satisfying these criteria were 
detected in each Taurus field. By plotting the difference between the observed instrumental magnitudes and those measured in 2MASS in one filter 
against an observed color (e.g., \mbox{$J_{O2K}-J_{2MASS}$} vs. $J_{O2K}-H_{O2K}$) the color correction terms between the {\sc Omega2000} and 2MASS filter systems were derived. This procedure was done for each observed Taurus field individually as observing conditions did vary between the individual fields. The mean transformation errors for each field and filter are also given in Table~\ref{o2kobservations}.


\subsection{{\sc SDSS} data} 
To differentiate between potential young members of Taurus and background and foreground contaminants/sources it is crucial to rely not only on NIR data but to take into account also optical information. The Sloan Digital Sky Survey (SDSS) offers up to five additional bands in the optical wavelength regime\footnote{http://photo.astro.princeton.edu/oriondatarelease/} \citep{Fin04} and covers parts of our NIR survey area. In the upper part of Table~\ref{o2kobservations} we provide estimates for the overlap between SDSS and our fields (in total $\sim$1.0 deg$^2$). For the rest of the paper we will only focus on the overlapping regions. We matched our {\sc Omega2000} NIR catalog with the SDSS data allowing a maximum offset of 2$''$ for an objects' position in both data sets. 

\subsection{{\sc Spitzer IRAC} data} 
Photometry at longer IR wavelengths allows us to detect the presence of disks around potential Taurus candidates supporting a young age of the object and thus membership of the SF region. Images taken with the IRAC instrument onboard the 
{\sc Spitzer Space Telescope} were downloaded from the data archive\footnote{These data are part of a large scale IRAC Taurus survey (PI: Deborah Padgett) and publicly available.} for the finally selected objects (see section~\ref{results}). The so-called "Post Basic Calibrated Data (PBCD)" were used to 
determine the flux of the individual candidates in the first two IRAC filters at 3.6 and 4.5$\,\mu$m. Unfortunately, the sensitivity of these images is insufficient to detect some of the fainter candidates at 5.8 and 8.0$\,\mu$m. The initial pixel values of the images 
were converted from MJy\,sr$^{-1}$ to count rates DN\,s$^{-1}$ (Data Numbers per second).
Afterwards, aperture photometry was carried out using the standard {\tt atv.pro}
routine in {\tt IDL} with an aperture size of 2 pixel and a sky annulus from 10-20 pixel. Aperture corrections were carried out as described in the IRAC Data Handbook to obtain the final magnitudes via $m =-2.5\,{\rm log}(x)+\Delta_{{\rm ZP}}$ with $x$ denoting the flux measured in DN\,s$^{-1}$ and $\Delta_{{\rm ZP}}$ being the zero point for each filter\footnote{Zero points taken from \citet{hartmann2005}: 19.66 (3.6$\,\mu$m), 18.94 (4.5$\,\mu$m), 16.88 (5.8$\,\mu$m), 17.39 (8$\,\mu$m).}. 

\subsection{{\sc VLT/ISAAC} data} 
For two of the three least luminous, and hence most interesting, objects we were granted Director's Discretionary Time (DDT) for ISAAC on ESO's VLT to obtain $J$-band low-resolution spectra. In low-resolution mode, ISAAC has a pixel scale of 0.147 $''$/pixel and a slit length 120$''$. The observations took place between February 7 and February 11, 2007.  The spectra cover a wavelength range between 1.1 and 1.4$\,\mu$m and we chose a slit width of 0.8$''$ resulting in a resolution of $R\approx680$. The observations were done in a nodding mode, with a 40$''$ nodding-throw along the slit. The detector integration time (DIT) was 120 s for both objects. The number of DIT exposures averaged out within the same frame (NDIT) and the number of frames (NINT) were 2 and 10, and 2 and 6 for CAHA Tau 1 and CAHA Tau 2, respectively. The airmass was between 1.5 and 2.0 in general, but increased up to 2.3 for the observations of CAHA Tau 2. The seeing was mostly better than 0.8$''$. The data reduction was done with the IRAF package {\tt onespec} following the standard procedure for low-resolution spectra: sky subtraction, flatfielding, cosmic ray removal, and finally extraction of the spectrum using an interactively fitted aperture. After wavelength calibration using argon lines the individual spectra for each object were averaged in combination with a 3-sigma clipping for the individual points. Telluric lines were corrected afterwards using standard stars\footnote{We used the B5 stars Hip023151 and Hip023946.} that were observed roughly at the same airmass directly after the science objects. To correct for the spectral response of the detector we compared the observed spectra of the standard stars with theoretical models of the same spectral type and applied correction factors to obtain an identical spectral slope.

\section{Results}\label{results}
\subsection{Candidate selection}
\begin{figure}
\epsscale{1.}
\plotone{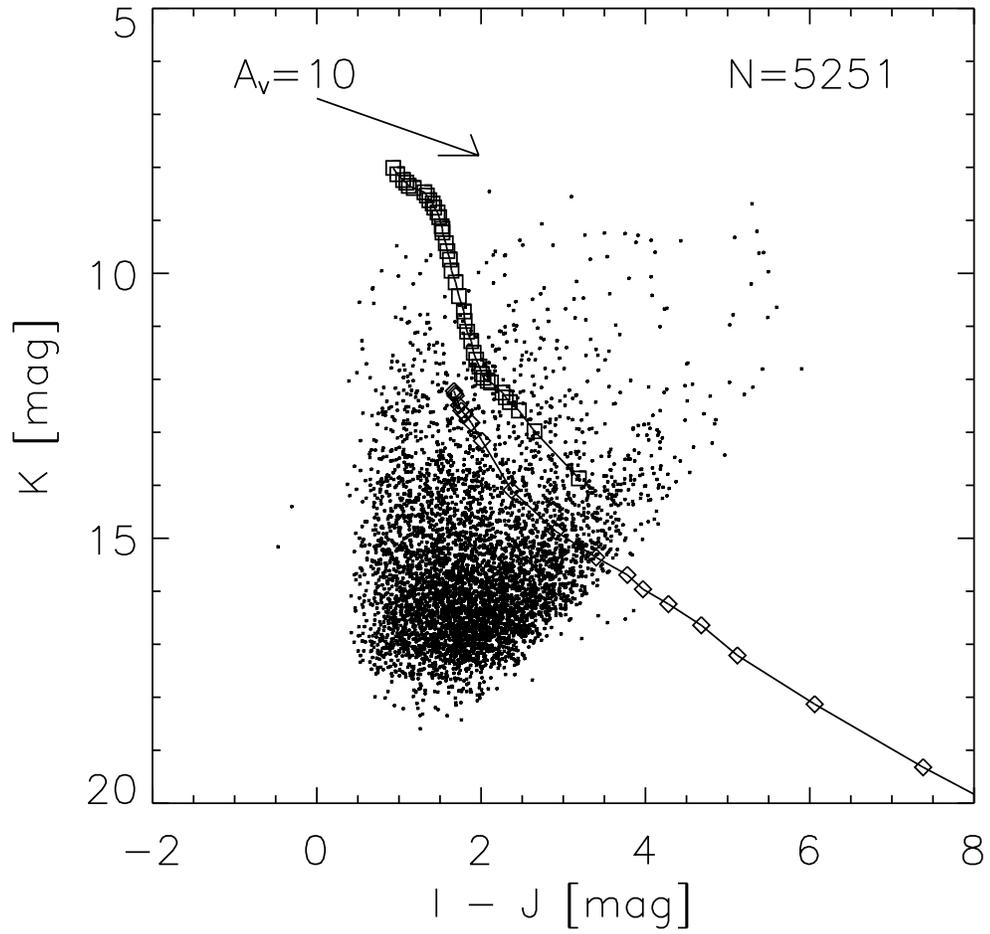}
\caption{Color-magnitude diagram based on optical (SDSS) and NIR ({\sc Omega2000}) data showing the 5251 objects fulfilling the selection criteria as described in the text. The NIR data is given in the CIT system, the optical filter is in the Johnson-Cousin system (see appendix). The solid lines with squares and diamonds are 5 Myr isochrones based on the models for low-mass stars and BDs from \citet{baraffe2002} and \citet{chabrier2000}, respectively. The extinction vector is based on the interstellar extinction law of \citet{mathis1990} with $R_V$=3.1. The distance to Taurus is assumed to be 140 pc.\label{selection1}}
\end{figure}

To identify potential low-mass objects of the Taurus region among the thousands of detected objects an optical/NIR color-magnitude diagram (CMD) served as starting point. In particular for fainter objects one has to find a way to distinguish between nearby low-mass objects, distant giant stars and also distant galaxies. While even distant galaxies with high red-shifts typically have  $I-J$ colors between 0 and 2.5, nearby low-mass objects normally appear significantly redder.
Figure~\ref{selection1}, where we plot K against $I-J$, takes advantage of this fact. As the overplotted theoretical isochrones (see below) are provided in the CIT photometric system, the NIR data in this figure were transformed into the CIT system. The optical data is in the Johnson-Cousin system. The required equations to transform between the different photometric systems are given in the appendix. In the plot we show only objects with a photometric error $<$\,0.1 mag in the NIR and with a S/N\,$>$\,5 in the SDSS $i$ and $z$ band. Out of more than 9000 objects that were commonly detected in the SDSS database and the NIR observations 5251 sources fulfilled these criteria. Theoretical isochrones computed by \citet{baraffe2002} and \citet{chabrier2000} are overplotted in the CMD. The isochrones have an age of 5 Myr, are scaled to a distance of 140 pc for Taurus \citep{kenyon1994} and cover the mass regimes between 0.02\,$M_\sun$ and 1.4\,$M_\sun$ \citep{baraffe2002}, and 0.002\,$M_\sun$ and 0.075$M_\sun$ \citep{chabrier2000}. The assumed age is a rather conservative assumption because, as mentioned in the introduction, Taurus seems to be significantly younger  \citep[$\tau\sim1$ Myr,][]{briceno2003}. Since younger objects tend to appear brighter (they have larger radii) and redder (e.g., due to circumstellar material), objects falling along the isochrones or lying above them are potential Taurus candidates and should be selected. As expected the least luminous objects are not selected as they appear too blue in  $I-J$ and are thus background stars or galaxies. In total, using the isochrones as a first selection criterion, the number of objects was reduced from 5251 down to 646.

\begin{figure}
\epsscale{1.}
\plotone{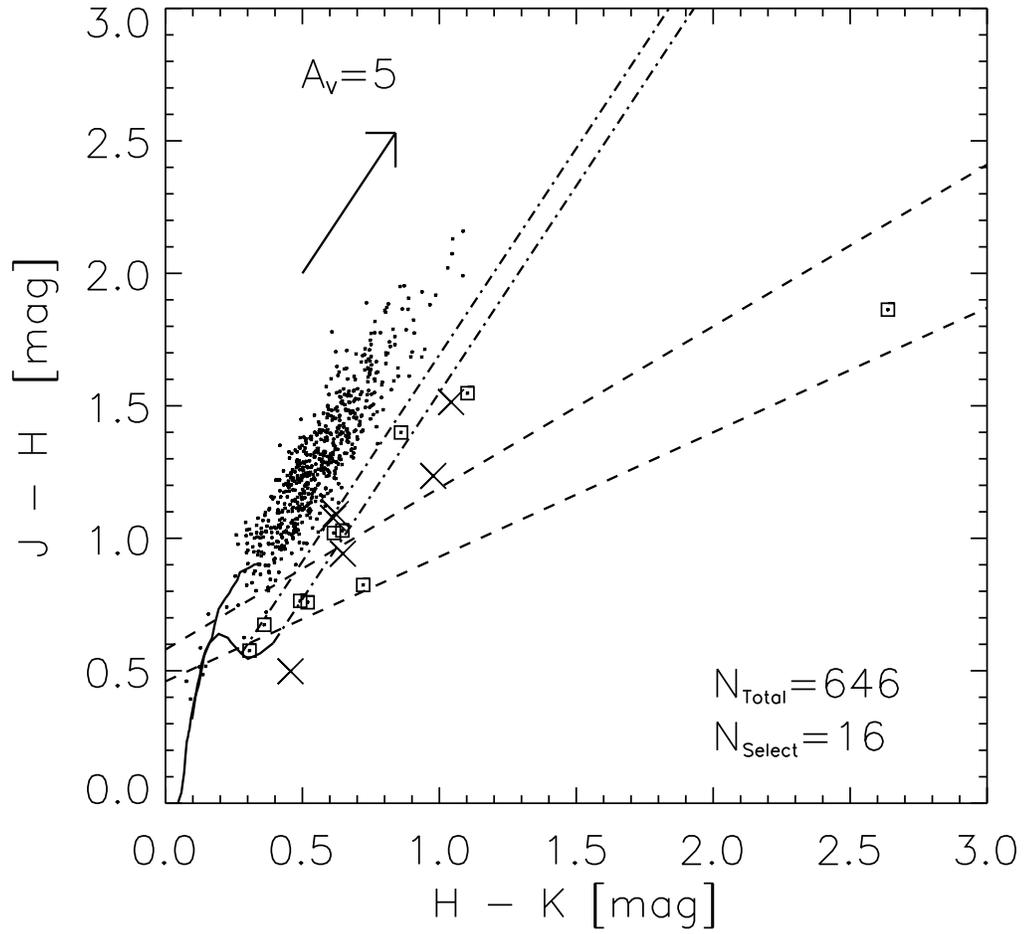}
\caption{NIR color-color diagram for the 646 selected objects (dots) and the finally selected 16 candidates. The boxes show the 10 already known members, while crosses indicate the positions of the final 6 low-mass candidates. The dashed lines denote the classical TTauri locus from \citet{meyer1997}, the solid lines show the main sequence stars from B8 to M6 and the class III giants from G0 to M7 \citep{besselbrett1988}. The dash-dotted lines show the directions in which an M6 and an M3 dwarf would be moved due to extinction. The extinction law is the same as in Figure~\ref{selection1}. The colors are given in the CIT photometric system.  
\label{selection2}}
\end{figure}

These objects are now, in a second step, plotted in a NIR color-color diagram (Figure~\ref{selection2}). Also here the CIT photometric system is used to be consistent with the first selection step. It can be seen that most of the selected objects lie along the reddened main sequence or along the reddened giant branch. Only very few sources populate the classical TTauri star locus \citep{meyer1997} or show otherwise any NIR excess emission. This indicates that the sample seems to be mostly contaminated by (background) stars. To minimize the possibility of selecting these objects, we considered only objects lying on the right hand side of a reddened M3 main sequence star and showing positive NIR colors. This selection resulted in 22 objects which were then subject to an individual inspection on the images. As some objects were too close to the edge of the detector in some filters and one object was observed twice in two adjacent telescope pointings the final sample of candidates consisted of 16 objects. Their positions in the color-color-diagram are indicated by crosses and squares in Figure~\ref{selection2}. 

Finally, the coordinates for the 16 remaining objects were crosschecked with the SIMBAD Astronomical Database\footnote{http://simbad.u-strasbg.fr/simbad/}. 10 objects were already listed (boxes in Figure~\ref{selection2}), but 6 candidates did initially not have a counterpart in the database (crosses in Figure~\ref{selection2}). 

\begin{figure}
\epsscale{1.}
\plotone{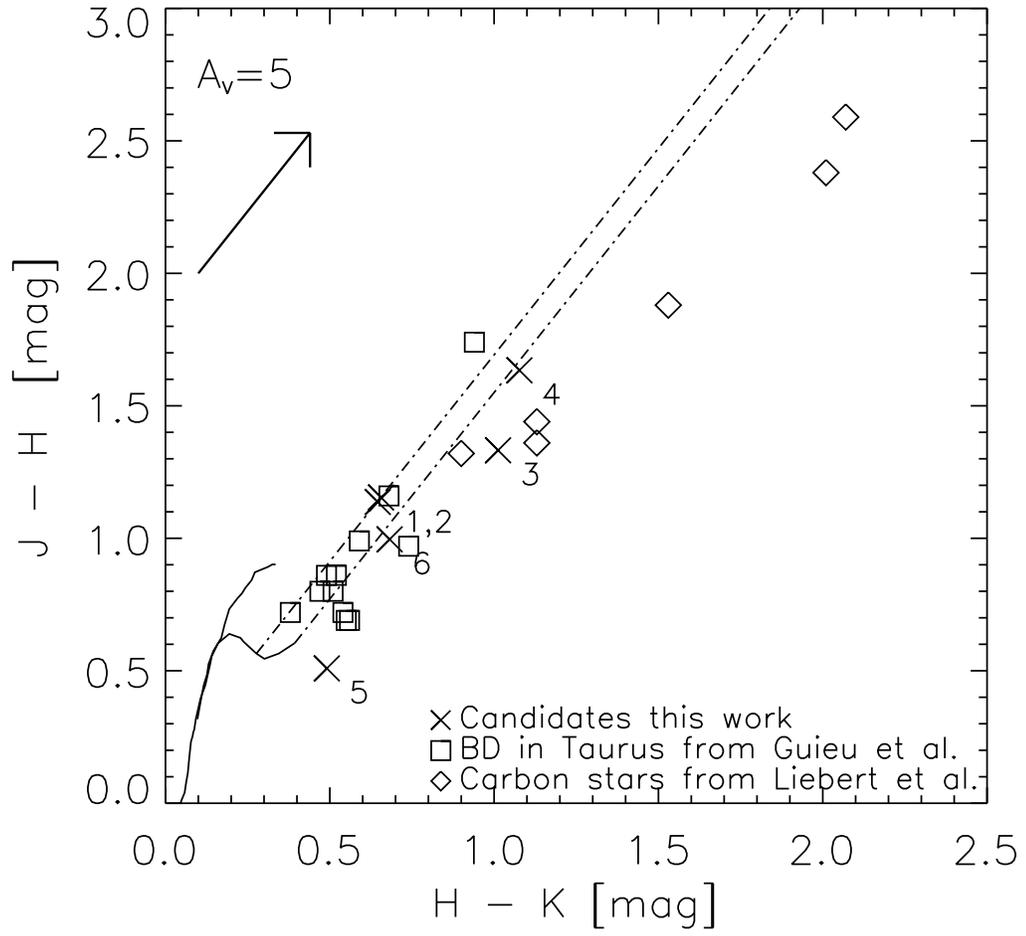}
\caption{NIR color-color diagram comparing the 6 new low-mass candidates presented in this paper (crosses), known Taurus BD from \citet{guieu2006} (boxes) and embedded carbon stars from \citet{liebert2000} (diamonds). The lines are the same as in Figure~\ref{selection2}. The colors are given in the 2MASS photometric system.\label{cc_nir}}
\end{figure}

For the 10 previously known objects the names, magnitudes, spectral types and references are summarized in Table~\ref{known_objects} along with our new NIR photometry. We stress that the objects with known spectral type are without exception low-mass stars and BDs in Taurus. This emphasizes the quality of our selection criteria. With KNPO-Tau 4 we even re-detected an object with a spectral type as late as M9.5. However, since we applied rather strict selection criteria, we did not re-identify \emph{all} of the known low-mass objects and BDs that lie in our observing fields. Some of them have for instance bluer $H-K$ colors than allowed by our selection criteria. One example can be seen in Figure~\ref{cc_nir}. In Table~\ref{known_objects} also the magnitudes from the 2MASS Point Source Catalogue are given for comparison. One can see that some objects appear quite variable while other sources apparently did not change in brightness at all. One possible explanation for the variability of the objects could be strong accretion events. CFHT-BD-Tau 19, for instance, shows clear evidence for accretion in its optical spectrum \citep{guieu2006}.

For the 6 objects that were initially not listed in SIMBAD the coordinates and photometric properties are shown in Table~\ref{candidates} together with estimates for their optical extinction based on extinction maps. However, during the revision process of this paper \citet{luhman2009caha6} also published results for our candidate CAHA Tau 6 and found the object to be very young BD with a spectral type of M9.25 and a mass of $\sim$0.015 M$_{\sun}$. Most interestingly, the object turned out to be a member of an isolated BD binary system with FU Tau (now FU Tau A) being the primary object which we also re-identified among the 10 previously known objects (see Table~\ref{known_objects}). Although CAHA Tau 6 (now FU Tau B) was thus already confirmed to be a young low-mass object, we keep it in our sample in the following, especially because no NIR data has been published so far. Thus, the properties of all our initially selected six candidates are discussed in greater detail in the following sections. NIR finding charts for CAHA Tau 1-5 are available from the online material. In \citet{luhman2009caha6} an optical image of the CAHA Tau 6 region is provided. 

\subsection{Photometric properties of candidates}\label{photometry}
In the following we will discuss the photometric properties our selected candidates. From now on we use the 2MASS photometric system in the NIR throughout the rest of the paper as most reference sources were published with 2MASS NIR data. In Figure~\ref{cc_nir} we show again a NIR color-color diagram. This time, however, we compare the colors of our six candidates to those of previously detected BDs in Taurus from \citet{guieu2006} and embedded carbon stars from \citet{liebert2000}. As we are interested in very low-mass objects we show only objects from \citet{guieu2006} with spectral types of M7 or later. We notice two things: First, the colors of most of our candidates do agree quite well with low-mass objects. Only CAHA Tau 5 appears a little too blue in $J-H$. Second, some of the carbon stars have NIR colors that are very similar to those of our young low-mass candidates. 
Especially CAHA Tau 3 (and to a lesser extend also CAHA Tau 4) finds itself surrounded by carbon stars.

\begin{figure}
\epsscale{1.}
\plottwo{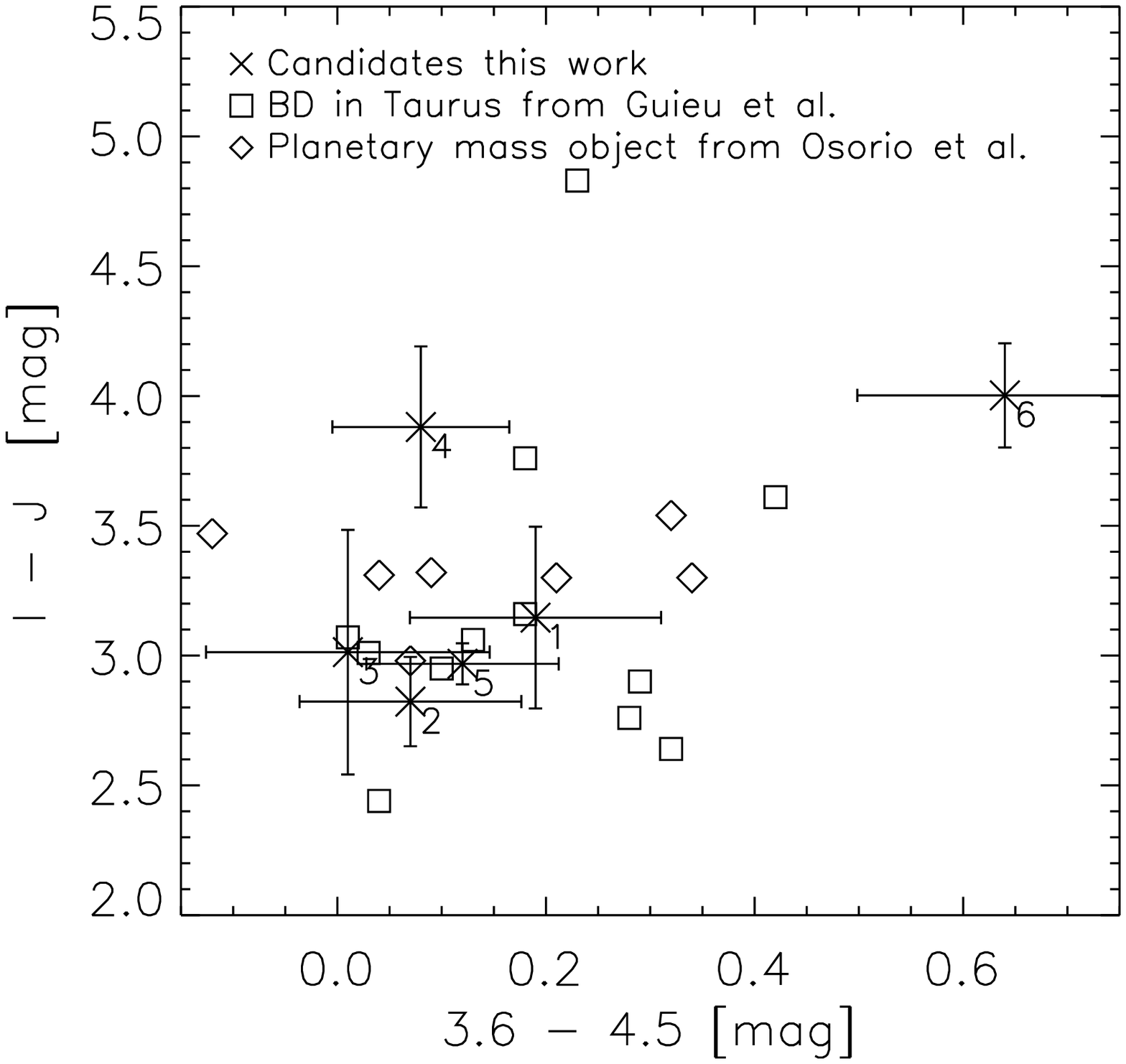}{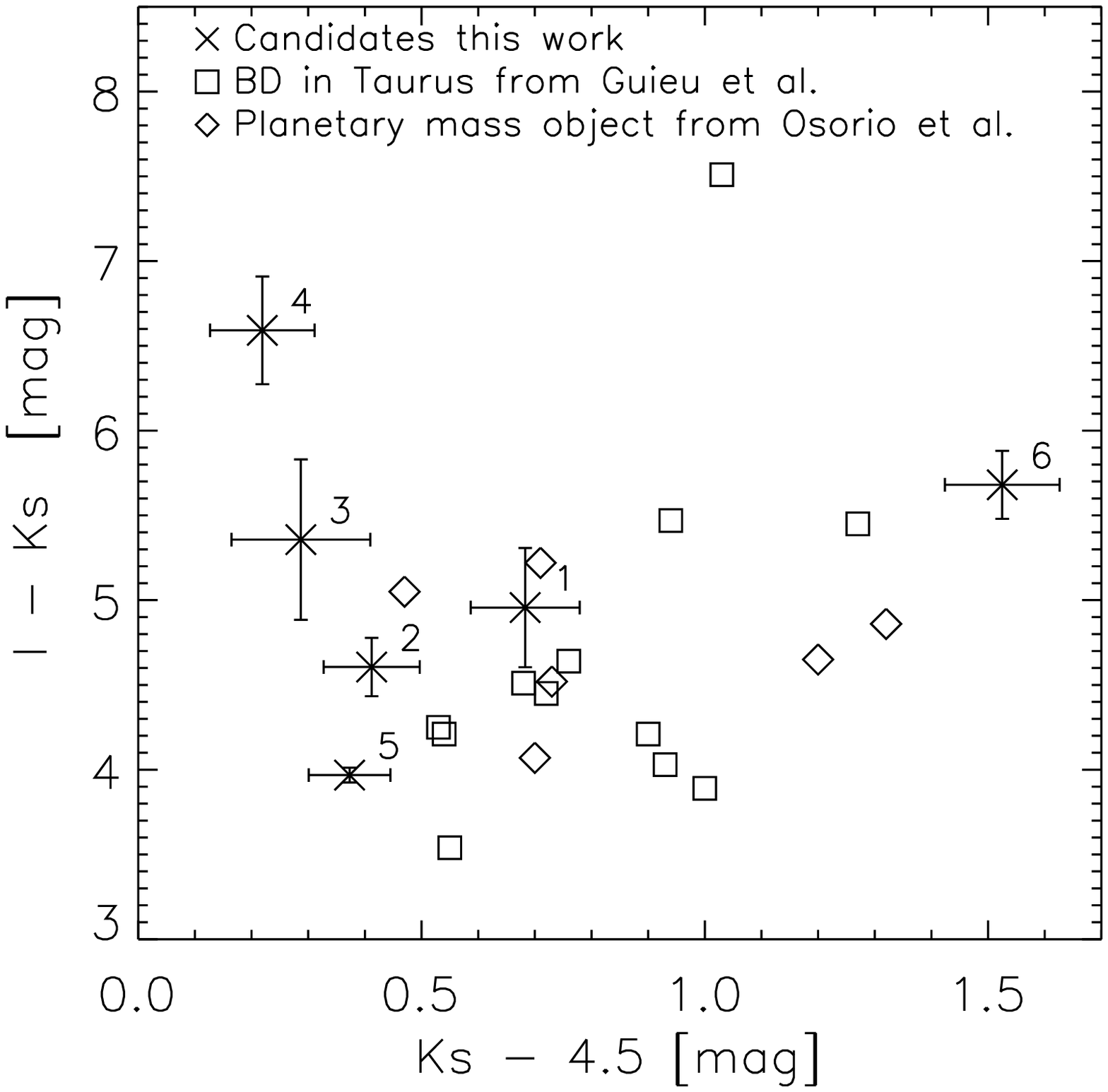}
\caption{Red/IR color-color diagrams comparing the 6 new low-mass candidates presented in this paper (crosses), known Taurus BDs from \citet{guieu2006} (boxes) and very low-mass BDs in the young $\sigma$ Orionis cluster from \citet{zapatero2007} (diamonds). The objects from Zapatero-Osorio typically have larger errors than our candidates. \label{cc_irac}}
\end{figure}

In Figure~\ref{cc_irac} we continue the comparison of our candidates to other young low-mass sources. This time, we increase the wavelength range in the NIR by using IRAC data at 3.6 and 4.5\,$\mu$m and we include optical colors in the diagrams. Furthermore, we add young, planetary mass candidates in $\sigma$ Orionis from \citet{zapatero2007} as reference objects. It shows that in the left-hand plot of Figure~\ref{cc_irac} our candidates populate nicely the same region of the color-color diagram as the other young low-mass objects. Only CAHA Tau 6 appears much redder in the IRAC color but \citet{luhman2009caha6} argued that this object is indeed particularly young ($\sim$ 1 Myr) and shows significant infrared emission excess. In the right-hand plot, where we have chosen a broader wavelength baseline for the colors, our candidates tend to occupy the left region of the plot, i.e., they are relatively blue in $K_s-4.5$\,$\mu$m compared to the reference objects. However, the colors are in general agreement with those of the other young low-mass objects. Only CAHA Tau 4 lies a little offsite compared to the other candidates and, again, CAHA Tau 6 appears very red in the $K_s-4.5$\,$\mu$m color.

\subsection{$J$-band spectroscopy}\label{spect}
As seen for instance in \citet{allers2006,allers2007} one has to be careful if one tries to derive stellar parameters of young objects such as spectral types, effective temperatures and masses solely from photometric measurements. Hence, we complemented our data with NIR $J$-band spectra for two of the three least luminous objects. We focused on these objects as they are potentially also the least massive and hence, for our purpose, most interesting candidates. 

\begin{figure}
\epsscale{1.}
\plottwo{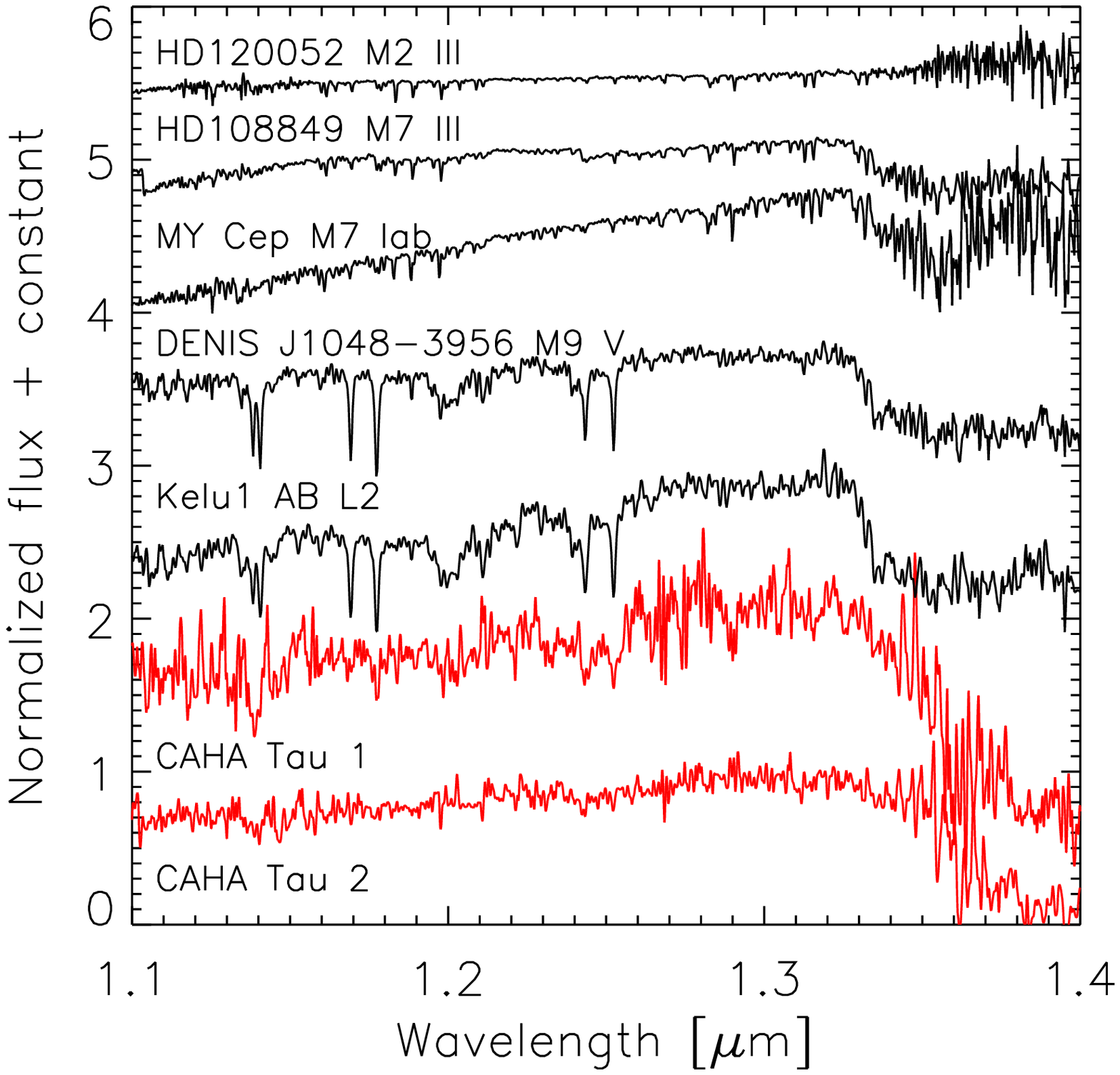}{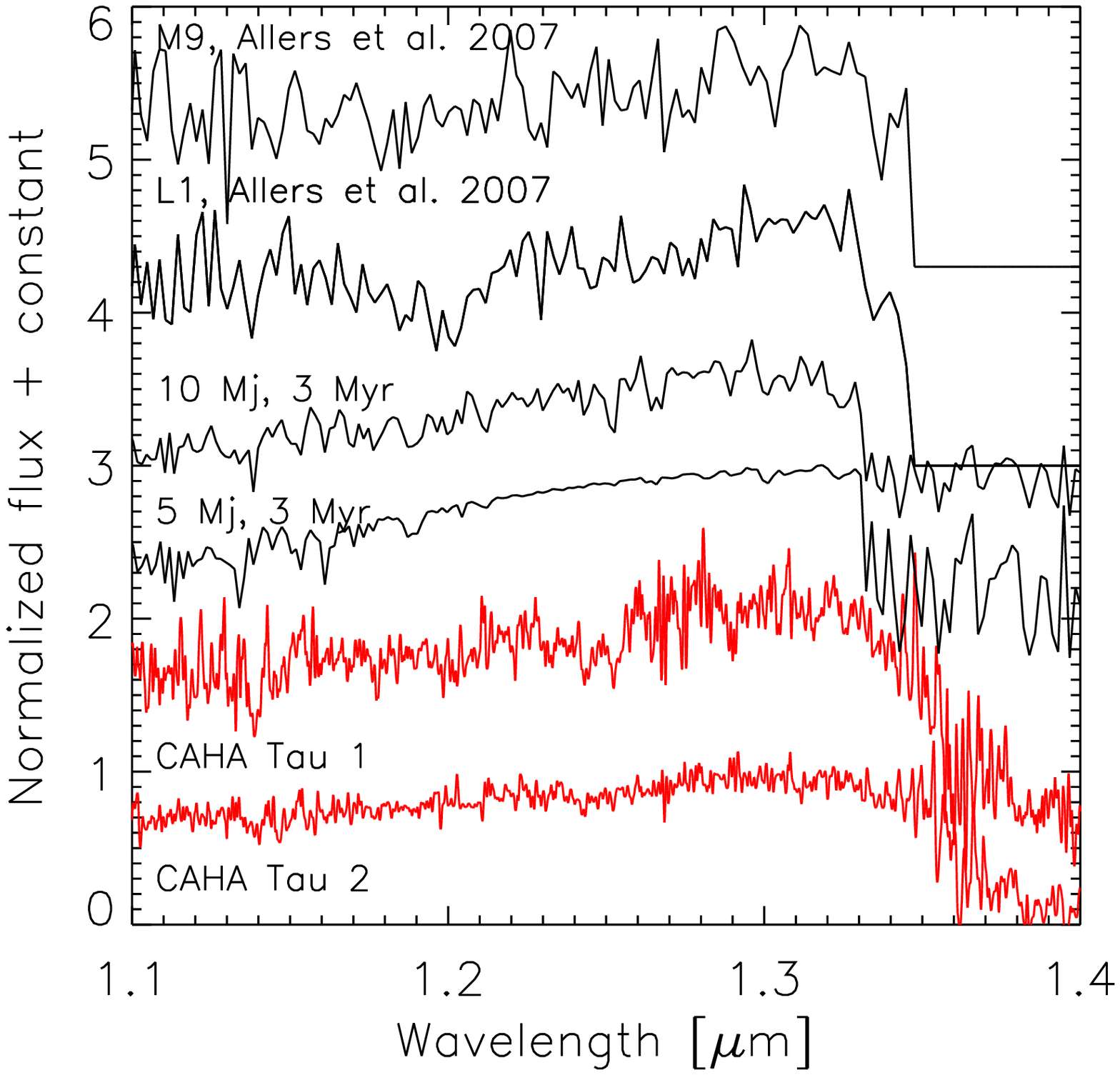}
\caption{\emph{Left:} $J$-band spectra of CAHA Tau 1 and CAHA Tau 2 compared to those of field BDs and M-type (super-)giants. The reference spectra have a resolution of $R\approx 2000$ and were downloaded from the IRTF spectral library \citep[http://irtfweb.ifa.hawaii.edu/$\sim$spex/IRTF\_Spectral\_Library;][]{cushing2005}. \emph{Right:} $J$-band spectra of CAHA Tau 1 and CAHA Tau 2 compared to theoretical models from \citet{burrows2003,burrows2006} with $R\approx 500$ and young low-mass objects from \citet{allers2007} with $R\approx 300$. Our candidate spectra were corrected with the $A_V$ values given in Table~\ref{candidates} and using the extinction law of \citet{mathis1990} with $R_V$=3.1 \label{spectra1}}
\end{figure}

\begin{figure}
\epsscale{1.}
\plotone{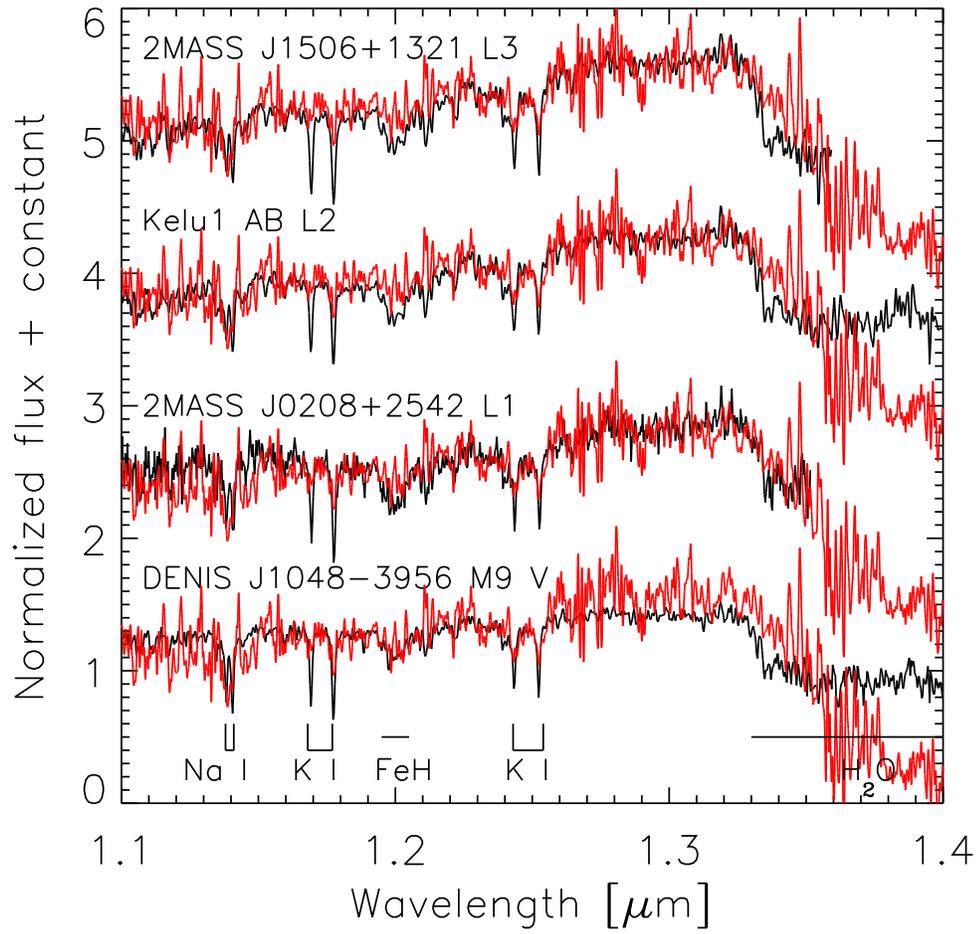}
\caption{Comparison of CAHA Tau 1 (shown 4 times in red) to low-mass field reference objects (black spectra) from \citet{cushing2005}. The best overall agreement is found with Kelu 1 AB a binary BD system where the combined spectrum has been established as reference for the spectral type L2. \label{spectra2}}
\end{figure}

In Figures~\ref{spectra1} we show the $J$-band spectra of CAHA Tau 1 and CAHA Tau 2 and compare them to data of field BDs and giants (left), and to young objects presented by \citet{allers2007} and model spectra from \citet{burrows2003,burrows2006} (right). The left hand side of Figure~\ref{spectra1} suggests that our candidates are late-type objects, as the spectrum of the early M-type giant HD 120052 does not show the deep water absorption band at the red edge of the spectrum that is seen in our candidates. Also, the general slope of the spectra is in agreement with late type objects. Furthermore it seems clear that our candidates have a low surface gravity, an indication for a possibly young age, as no strong spectroscopic features are present. However, in the spectrum of CAHA Tau 1 we find weak spectral signatures: There is a dip near the position of the Na I doublet around 1.14\,$\mu$m and also hints for the K I doublets near 1.17 - 1.18\,$\mu$m and 1.24 - 1.25\,$\mu$m which show up much stronger in the spectra of the higher surface gravity field stars. In Figure~\ref{spectra2} a direct comparison between the $J$-band spectrum of CAHA Tau 1 and those of late type field objects is shown. The overall very similar shape and slope is striking and also the existence of similar spectral features, even if they are very weak in case of CAHA Tau 1. The best overall agreement in shape and slope between a reference spectrum and CAHA Tau 1 is found for Kelu 1 AB. Based on this we estimate the spectral type of CAHA Tau 1 to be L2$\pm$0.5 which is the latest spectral type known in Taurus today. We will discuss this finding in more detail in section~\ref{caha1}. 

A spectral classification for CAHA Tau 2 is difficult as its spectrum lacks any spectral lines. However, it is not unusual for young low-mass sources to have almost no spectroscopic features as can be seen in \citet{mcgovern2004} and \citet{lodieu2008}. The first group published spectra of the young Taurus object KPNO-Tau 4 and of $\sigma$ Ori 51 while the second group presented spectra of young L dwarfs in Upper Sco. All of these objects showed barely any spectral features in the $J$-band, presumably due to their low surface gravity, and appear in this respect similar to  CAHA Tau 2\footnote{As mentioned above we re-identified KPNO-Tau 4, a confirmed young and low-mass brown dwarf, during our selection process (see Table~\ref{known_objects}).}. 
Finally, a comparison to theoretical models and reference spectra of young low mass objects further emphasizes the almost featureless behavior of these objects (right hand side of Figure~\ref{spectra1}). 

One thing, however, appears strange in the spectrum of CAHA Tau 2 and so far we lack a satisfying explanation if it was a young object. The drop in the flux caused by water vapor around 1.33\,$\mu$m is not as steep as in the other low-mass objects including CAHA Tau 1. Although the peak flux in the $J$-band is reached between 1.32 and 1.33\,$\mu$m, as it is also the case for the reference BDs, the flux of our candidate decreases more slowly with increasing wavelength. This behavior is not typical for young, low-mass objects \citep[see, e.g.,][]{lodieu2008}. However, $J$-band spectra of C-N and C-J carbon stars as presented by \citet{tanaka2007} show a similar behavior with no significant drop in the flux up to 1.35\,$\mu$m. And also the flux of the M7 III star HD108849 in Figure~\ref{spectra1} shows a rather modest decline longward of 1.33\,$\mu$m in addition to an overall featureless $J$-band spectrum. 

To summarize, for CAHA Tau 1 we find spectroscopic support for the young and low-mass nature of the object from both, the weak spectral features and the overall slope of the spectrum, and we can estimate a spectral type. For CAHA Tau 2 a definite spectroscopic proof that the object is indeed a young, low-mass Taurus member is pending and although the general slope is very similar to that of CAHA Tau 1 the lack of strong water absorption around 1.33\,$\mu$m talks in favor of a background carbon star or giant.

\subsection{Proper motion of Taurus members and our candidates}

Another way to find additional support for the low-mass nature of our candidates is to analyze their proper motion. For CAHA Tau 6 \citet{luhman2009caha6} already showed that it most likely shares common proper motion with FU Tau A which itself shows a proper motion consistent with Taurus membership. Thus we focus on the remaining 5 objects.

\begin{figure}
\epsscale{1.}
\plotone{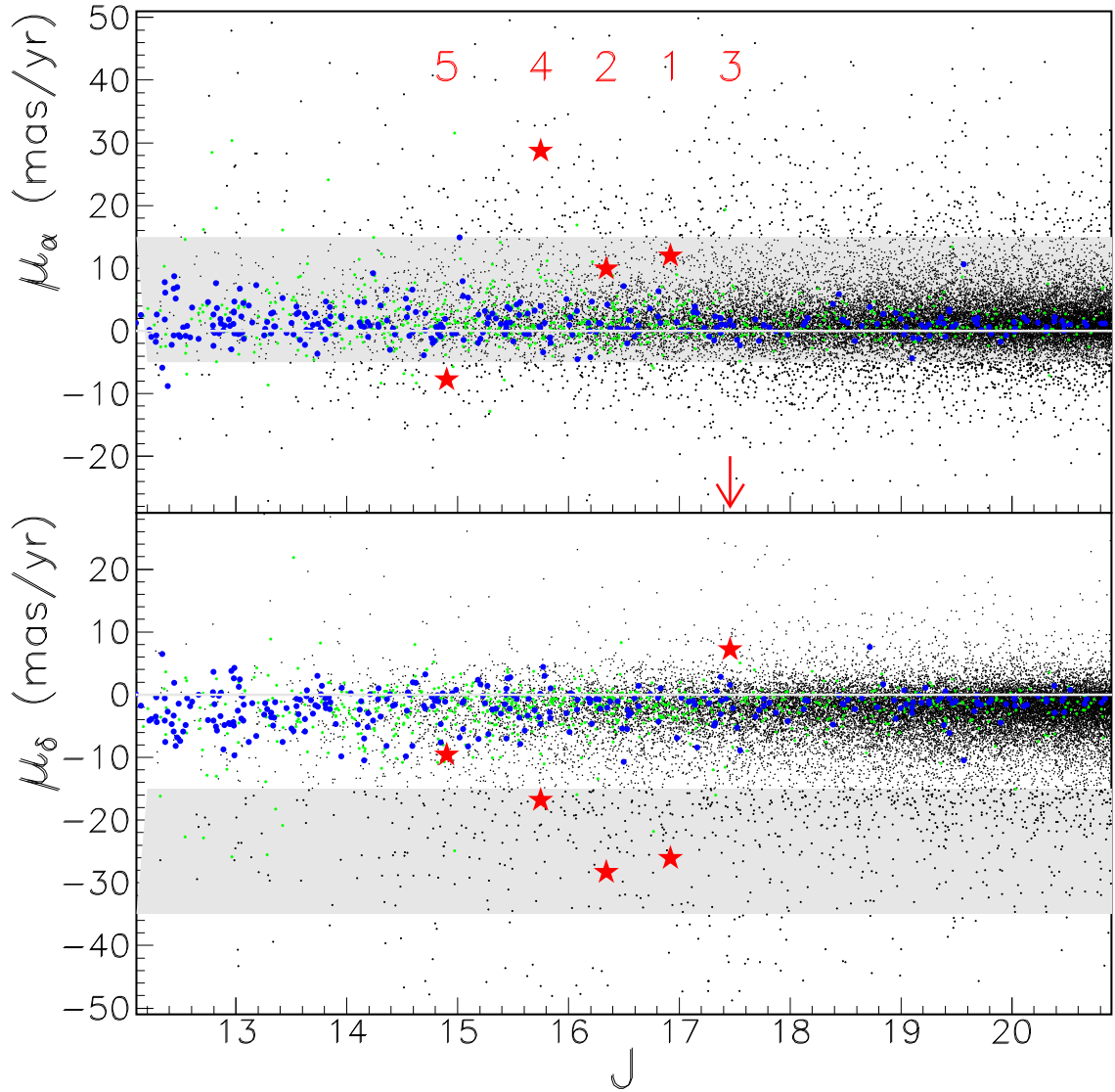}
\caption{Absolute proper motions based on a simulation using the Besan\c{c}on model.
Black dots are dwarfs, larger green dots are sub-giants, largest blue dots are giants.
The candidates' relative proper motions are indicated. 
The grey area gives the proper motion locus of most Taurus members (see also Figure~\ref{PMcaha}).
\label{PMsim}}
\end{figure}

Proper motion studies of stellar members of the whole Taurus-Auriga association show a wide dispersion in proper motions along the right ascension axis, with \mbox{$-5<\mu_\alpha<15$\,mas/yr}, but a consistently negative proper motion along the declination axis, with \mbox{ $-35<\mu_\delta<-5$\,mas/yr} \citep{Fri97}, mostly due to the reflex motion \citep{Jon79}. 
We have run a simulation of our fields using the Besan\c{c}on model \citep{Rob03,Rob04}, adding an extinction layer at 150\,pc distance with variable extinctions. 
The model does not include the star forming region, only the smooth components of the Galactic disks and halo. 
  This simulation (see Figure~\ref{PMsim}) shows that most giant and sub-giant stars have proper motion $|\mu|<5$\,mas/yr, corresponding to halo-type velocities (at most) and distances larger than 10\,kpc (see subsection\,\ref{carbon}). The fraction of giant and sub-giant stars with proper motion $\mu_\delta<-10$\,mas/yr and $16<J<18$ is $3\pm1\%$ (1-$\sigma$ statistical error). The fraction for the brighter M~giants would be even smaller. 
  Therefore, a significantly larger proper motion measurement would reject the giant hypothesis, while a proper motion $\mu_\delta>-8$\,mas/yr or $|\mu_\alpha|>30$\,mas/yr would most likely exclude a Taurus membership. Because the Taurus proper motion space overlaps with that of disk dwarfs, proper motions cannot completely remove the field dwarf contamination.  A fraction of 5\% of field dwarfs with $16<J<18$ have a proper motion of \mbox{$|\mu_\alpha|<30\,$mas/yr} and \mbox{$\mu_\delta<-15$\,mas/yr}.
  
\begin{figure}
\epsscale{1.}
\plotone{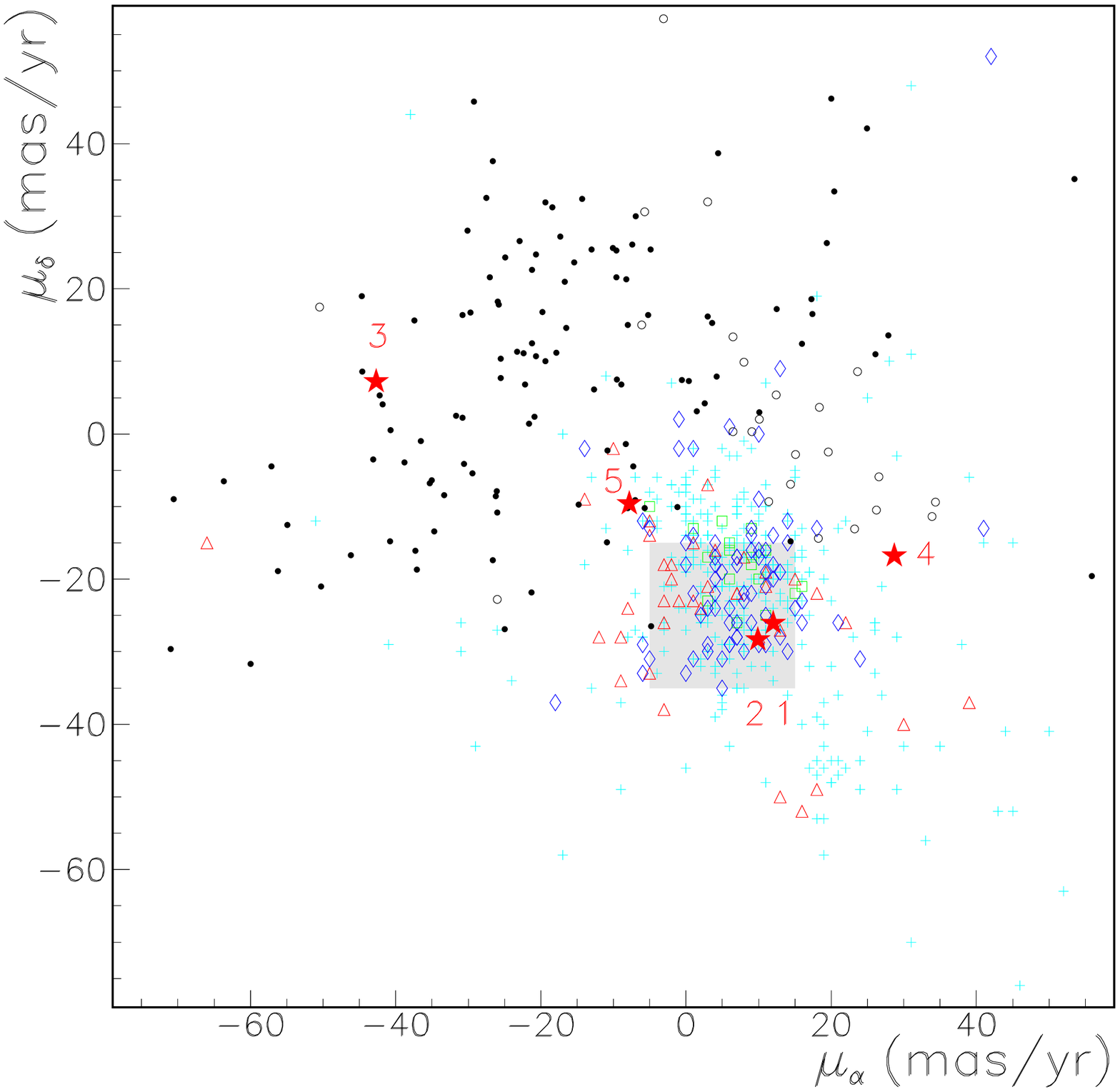}
\caption{Proper motions of our candidates compared to proper motions of Taurus members from the literature: the compilation of \citet{Duc05} (pale blue pluses), \citet{Har91} (green squares, objects with errors smaller than 10\,mas/yr in both directions), \citet{Fri97} (red triangles) and \citet{Jon79} (blue diamonds). 
Typical errors for the two latter references are 5\,mas/yr. Also shown are the proper motions of stars of similar magnitudes as our two candidates CAHA Tau 1 (thick dots) and CAHA Tau 2 (empty circles).
\label{PMcaha}}
\end{figure}

   To complement our near-infrared and SDSS images, we used public data from CFHT's 12k and MegaCam cameras, and, for CAHA Tau 1, the catalogues from the UKIDSS Galactic Plan Survey $K$ band data from the Data Release~3 \citep{Law07}. 
   CFHT images were detrended using the Elixir pipeline \citep{Mag04}. 
   Object detection and astrometry was performed using {\tt SExtractor} \citep{Ber96} on all images (CFHT, SDSS, {\sc Omega2000}). 
   Bright objects (SNR$>$100, a few dozens) were used to derive a preliminary transformation, which is refined with all objects with SNR$>$10 (one hundred). 
   For the three faintest targets, 
   2MASS data have too large positional uncertainties to improve the fit. 
   The photonoise errors are negligible at our targets' brightness; the error budget is dominated by transformation errors, which includes refraction, instrumental distortions and proper motions of the reference stars. Therefore we set the positional accuracies to a minimum of 30\,mas, typical of the best position dispersion measured by {\tt scamp} for bright stars. 
   This dispersion rises to 80\,mas in some cases (e.g. when using SDSS as the reference catalogue). 
   We then fit a proper motion on the CFHT, SDSS, O2000 and UKIDSS positions when a detection was found within 1\,arcsec of the O2000 position, for all objects with SDSS~$z$ and O2000~$K_s$ magnitudes similar to our candidates (see Table\,\ref{PMcand} and Figure~\ref{PMcaha}).
   We have not corrected for the velocities of the stars which form our reference grid and we present relative proper motions only. The Besan\c{c}on  simulation shows that stars brighter than $J<16$\,mag have a small averaged proper motion of $\langle\mu_\alpha\rangle=2$\,mas/yr and $\langle\mu_\delta\rangle=-4$\,mas/yr, although some have large proper motions.

  The data set used for these measurements is far from optimal: we compare positions obtained with four different instruments, with the first epochs obtained in the red and the last epoch(s) obtained in the near-infrared. We did not find any correlation between positional shifts and hour angles, or filter bands, so that most of the transformation error budget is likely to be due to the imperfect corrections of the (relative) instrumental optical distortions, and the proper motion dispersions of the reference objects. Because we could not calculate a global astrometric solution simultaneously for all instruments, proper motion measurements could not be determined during the registration process itself.
      
\subsubsection{CAHA Tau 1 and 2}

   Despite these limitations we find that the proper motions of CAHA Tau 1 and CAHA Tau 2 are compatible with typical Taurus members' proper motions (Figures~\ref{PMsim}, \ref{PMcaha}). In order to judge whether the giant hypothesis is rejected, we need to know how significantly they differ from $\mu=5$\,mas/yr.
  Because of the short time baseline and the required proper motion accuracy, it is important to have a good estimate on the proper motion errors, which will be smaller than the signal by a factor of a few at most.
  A first estimate of the positional accuracy of each data set is given by the shift dispersion of the bright reference stars, subtracting the estimated reference catalogue errors, to which we add the photonoise of the candidate. 
  However, this might be a lower limit for CAHA Tau 1 as it is located in a corner of the CFHT CCDs, in an empty area. The location of CAHA Tau 2 
  is more central and favorable. The colors of the candidates are typical of the bulk of the reference stars. Many reference stars are background stars with significant reddening, and therefore red colors. Furthermore, the difference in colors does not correlate with proper motions or dispersions around the proper motion fit.
  The large number of CFHT images at a given epoch for CAHA Tau 1 allows us to measure the positional dispersion. As described above, we used this value as a proxy for the positional accuracy. Unfortunately, this robust method cannot be used for {\sc Omega2000}, and even less for the UKIDSS single image. 
  Moreover, the UKIDSS DR3 catalogue comes with no astrometric errors, and the proper motion fit relies crucially on this single point, as it extends by one~year the time baseline.
%
 We derive the proper motion error from the dispersion of the proper motions of objects of similar magnitudes as our targets. We correct for the intrinsic dispersion based on the Besan\c{c}on model simulation, although we estimate that measurement errors are the main contributor to the total dispersion. These two arguments suggest that our error estimates are correct within a few mas/yr.
  
In that case, the proper motion of CAHA Tau 1 is incompatible with a giant star proper motion at the 1.3$\sigma$ level; CAHA Tau 2 at the 1.9$\sigma$ level.
We note that the fit of CAHA Tau 2 proper motion along the right ascension is not good, and a value larger than Taurus members' proper motions cannot be confidently excluded. 

\subsubsection{CAHA Tau 3, 4 and 5}  

Unfortunately the situation for CAHA Tau 3, CAHA Tau 4 and CAHA Tau 5 is different. The data set for these targets is more limited than for CAHA Tau 1 and CAHA Tau 2, in particular no UKIDSS data are available. Also the positional dispersions of reference stars is larger, pointing to a poorer registration. Based on the current data we find that the proper motions we measure for CAHA~Tau~3 do not correspond to the typical proper motions of known Taurus members (see Table~\ref{PMcand} and Figure~\ref{PMsim}). 
However, the current error bars are large and our measurements are not stable against the registering parameters' fine tuning, so that further observations are required to ensure a robust measurement. The proper motions of CAHA~Tau~4 and CAHA~Tau~5
are in agreement with a Taurus membership but have little contamination rejection power.

\section{Discussion}\label{discussion}

\subsection{CAHA Tau 1 \--- the first L2 dwarf in Taurus}\label{caha1}
From the results shown above we believe that CAHA Tau 1 is a young, low-mass object in Taurus with spectral type L2$\pm0.5$ and thus so far the object with the latest spectral type in Taurus \citep{luhman2009xray}. To estimate its effective temperature we used the relation between spectral type and $T_{eff}$ for low-mass field objects found by \citet{golimowski2004}. Based on the same reference we also derived the bolometric correction in the $K$-band. All values are summarized in Table~\ref{caha1_table}. The errors are based on the uncertainties in the spectral type and the intrinsic errors of the conversions found by \citet{golimowski2004}. 

\begin{figure}
\epsscale{1.}
\plotone{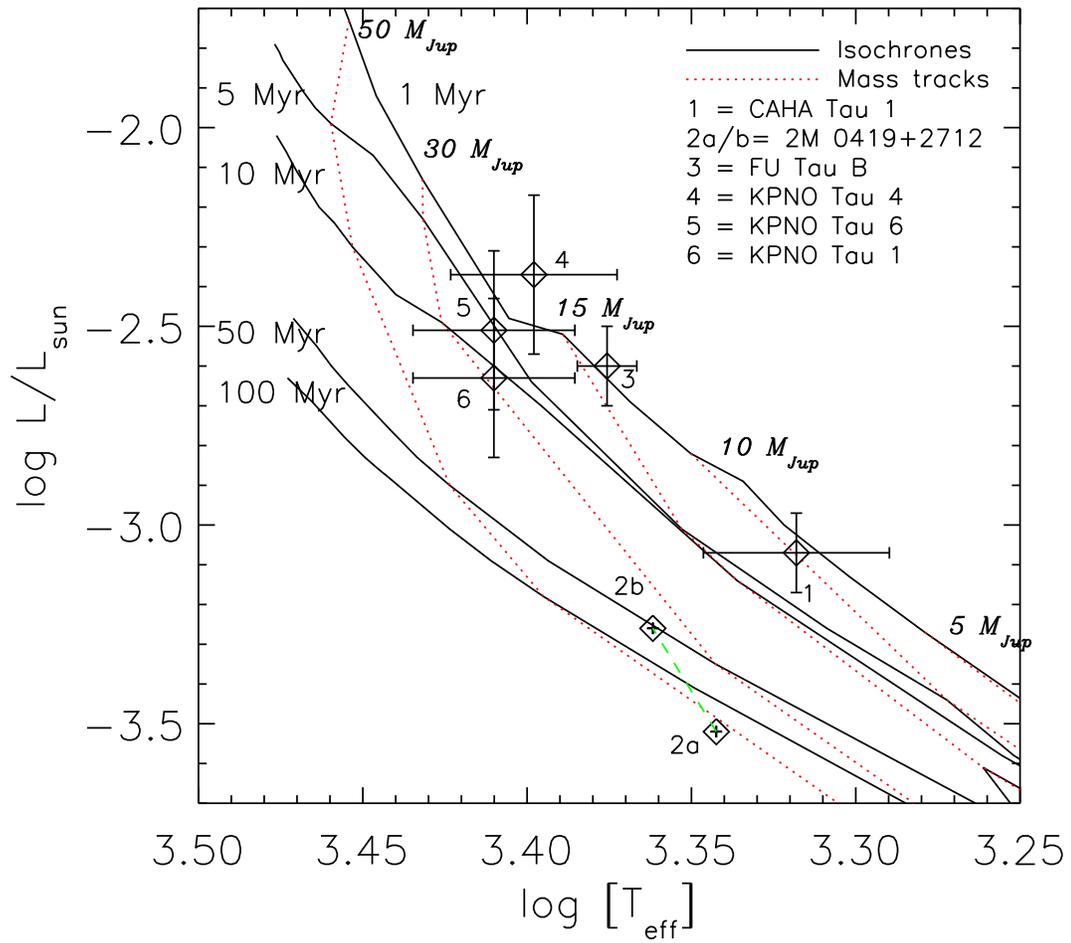}
\caption{HR-diagram comparing some of the least massive Taurus members known today.\label{HR-diagram}}
\end{figure}

In Figure~\ref{HR-diagram} we plot an HR-diagram for CAHA Tau 1 and known low-mass Taurus objects with spectral types $\ge$M9.0. The luminosity of CAHA Tau 1 is derived from the bolometric correction $BC_K$ (Table~\ref{caha1_table}) and assuming a distance to Taurus of 140 pc and $A_V$=2.3. The reference objects were collected from \citet{luhman2009caha6,luhman2009xray} and \citet{guieu2006}. Overplotted in the diagram are isochrones (1, 5, 10, 50, and 100 Myr) and mass tracks (5, 10, 15, 30, and 50 M$_{Jup}$) from \citet{chabrier2000}. Three points are noteworthy: (1) For most objects the derived properties are in good agreement with young ($\le$ 10 Myr), low mass ($\le$ 30 M$_{Jup}$) objects. (2) CAHA Tau 2 (number 1 in the plot) falls nicely between the 1 and 5 Myr isochrones and, depending on its actual age, its observed properties translate into a most likely mass between 5 and 15  M$_{Jup}$ based on theoretical models (Table~\ref{caha1_table}). It is thus the coolest Taurus member known today and only the second object in the region with a mass most likely below the Deuterium burning limit making it a free floating planetary mass object. (3) The other object that was recently identified as a Taurus member of planetary mass with only 4-7 M$_{Jup}$ \citep[][2MASS 0419+2712, number 2a in the plot]{luhman2009xray} lies below the 100 Myr isochrone and close to the 50 M$_{Jup}$ mass track. This behavior was already discussed in the discovery paper where the authors concluded that the old isochronical age for the object is likely a reflection of errors in the adopted temperature scale and evolutionary tracks. Similarly, the coolest member of the Chameleon I SF region (spectral L0) appears also too old for its luminosity \citep{luhman2008chameleon} which is comparable to that of 2MASS 0419+2712. Disregarding the derived temperature and assuming an age of 1-3 Myr the derived luminosity of  2MASS 0419+2712 yields then a mass of 4-7 M$_{Jup}$ as mentioned above. While there are certainly notable uncertainties in the temperature scales and evolutionary models, it is interesting to note that all other late type Taurus members shown in Figure~\ref{HR-diagram} (incl. CAHA Tau 1) populate as expected the young areas of the diagram. Thus, it is worth investigating possible astrophysical reasons why 2MASS 0419+2712, if it is indeed a Taurus member, should appear underluminous. As a first test, we use the same temperature scale and bolometric correction that we used for CAHA Tau 1. This leads indeed to a significant shift in the HR-diagram (number 2b): Instead the of original values of $T_{eff}$=2200 K and log L/L$_\sun\approx$3.52, we find  $T_{eff}$=2300 K and log L/L$_\sun\approx$3.26, demonstrating the uncertainties across the different temperature scales. However, 2MASS 0419+2712 appears still too low in the diagram compared to the other sources. As a second source of uncertainty, we speculate that part of this might result from underestimating the extinction towards the source. Even if no disk, and in particular no edge-on disk creating significant extinction, appears to be present \citep{luhman2009xray}, the assumed value of $A_J$=0 mag, hence no extinction, for 2MASS 0419+2712 might be too low. We note that also in the case of the L0 dwarf in Chameleon I \citep{luhman2008chameleon} no extinction was assumed. Finally, we mention that it could still be possible that 2MASS 0419+2712 is indeed older than typical Taurus members or it is not associated with the main Taurus population and further away than the assumed 140 pc. In all these case the mass of the object would be higher than the current estimates of a few Jupiter masses. More spectroscopic data and a proper motion analysis might yield additional insight into this puzzle. 

\begin{figure}
\epsscale{1.}
\plotone{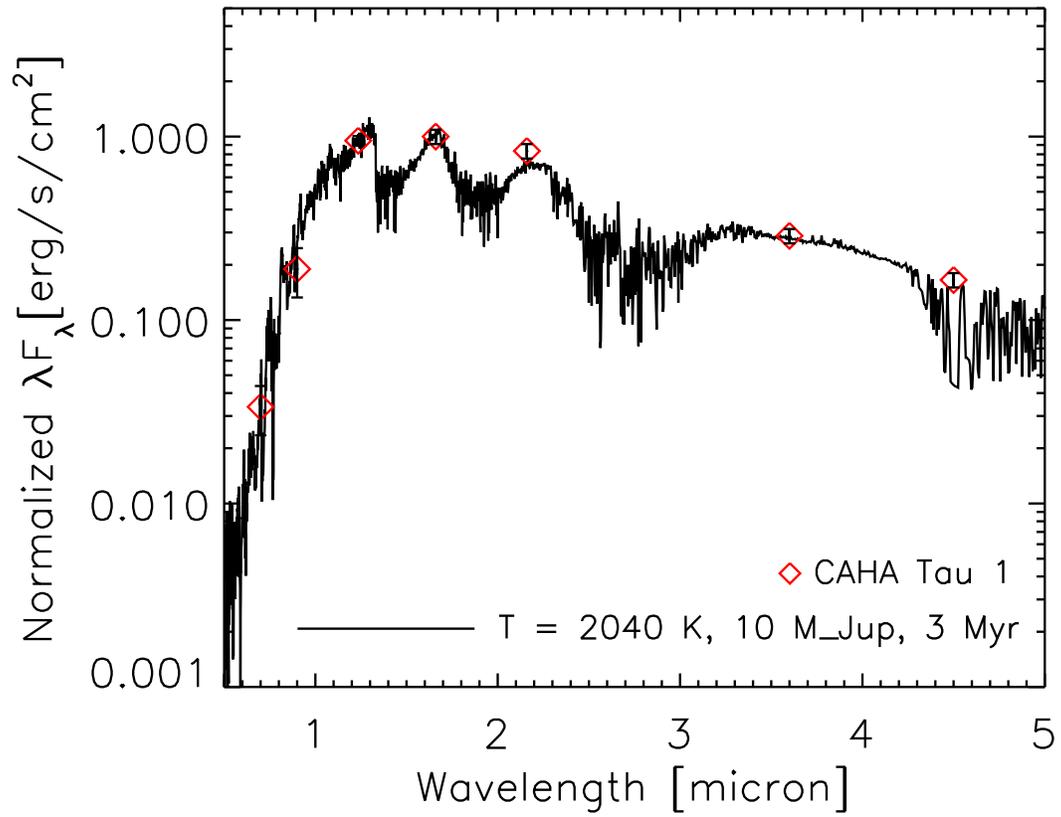}
\caption{Broad band SED of CAHA Tau 1 overplotted on theoretical model for a 3 Myr 10 Jupiter mass object with T$_{eff}$=2040 K  \citep{burrows2003,burrows2006}. \label{caha1sed}}
\end{figure}

As CAHA Tau 1 fits nicely on the HR diagram we compare its broad band SED with simulations. In Figure~\ref{caha1sed} we plot the dereddened photometry together with a theoretical model SED for a 3 Myr object of 10 Jupiter masses with $T_{eff}$=2040 K \citep{burrows2003,burrows2006}. The fluxes were normalized to unity in the $H$-band. The observed photometry was corrected for extinction using the value for $A_V$ given in Table~\ref{candidates} and the extinction curve from \citet{mathis1990} with $R_V$=3.1. The agreement between the model and the observations is very good and only the observed K-band and 4.5\,$\mu$m fluxes appear a little too high compared to the simulations. At least for the flux at 4.5\,$\mu$m it might well be that circum(sub-)stellar material creates  excess emission above the photosphere. \citet{zapatero2007} found also excess emission for their young, very low-mass objects but mostly longward of 5\,$\mu$m. It would be nice to check whether the apparent emission excess we see in CAHA Tau 1 extends also to longer wavelengths.


\subsection{CAHA Tau 2 - 5 \--- Embedded carbon stars or (super-)giants?}\label{carbon}
\begin{figure}
\epsscale{1.}
\plotone{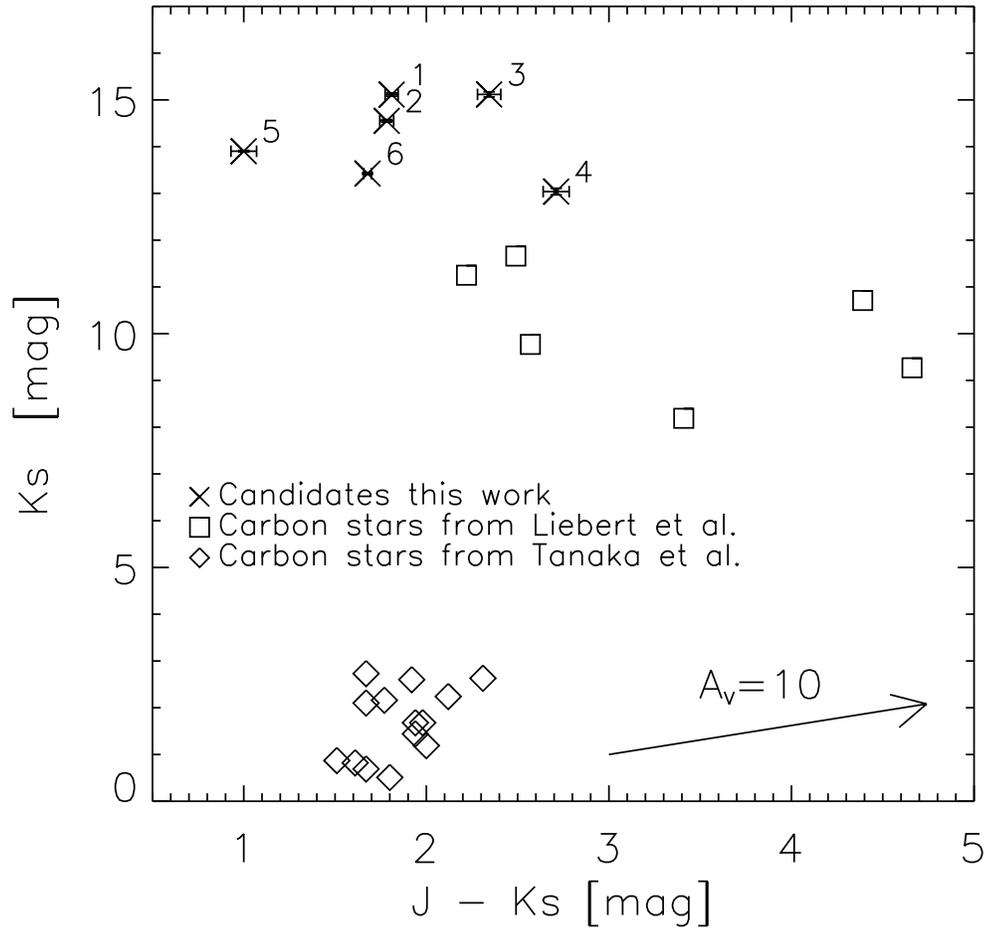}
\caption{NIR color-magnitude diagram (in the 2MASS system) comparing the 6 new low-mass candidates presented in this paper (crosses) with two sets of carbon stars. The objects from \citet{liebert2000} lie outside the galactic plane, have assumed distances between $\sim$9 and $\sim$110 kpc and are embedded in dusty shells. Nearby carbon stars of types C-J and C-N from \citet{tanaka2007} are shown with diamonds.\label{cm_carbon}}
\end{figure}

While for CAHA Tau 1 there are sufficient indications that it is a young Taurus member, it has not yet been convincingly shown that our candidates 2 - 5 are \emph{not} background carbon stars or late-type (super-)giants. And eventually only additional spectra will help to settle this open question. However, to elaborate a little further on this problem, we plot in Figure~\ref{cm_carbon} a color-magnitude diagram with our candidates and the two groups of carbon stars we already mentioned above \citep{tanaka2007,liebert2000}. For completeness also CAHA Tau 1 and CAHA Tau 6 (alias FU Tau B) are shown in the plot. One finds that the very distant and deeply embedded objects from \citet{liebert2000} tend to be much redder than our candidates while the nearby carbon stars from \citet{tanaka2007} have a very similar $J-K_s$ color like the bulk of our objects. Even more important, however, is the fact that all of our candidates appear fainter than even the most distant objects from \citet{liebert2000} which are already assumed to lie in or even beyond the Galactic halo. Hence, if our candidates were distant objects then (1) they should appear redder than they are as the Taurus molecular cloud should cause significant additional extinction and (2) they should be extremely far away. In Table~\ref{giant_magnitudes} we list the typical values for absolute K-band magnitudes of carbon stars and M-type giants. Based on these figures and the apparent magnitudes of our remaining 2 least luminous candidates (CAHA Tau 2 and 3; Table~\ref{candidates}) we derive a range for the objects' potential distance moduli ($m_K-M_K$). We focus again on these objects as they are intrinsically more interesting due to their faintness. Using \begin{equation}
d=10^{\,0.2\cdot(m_K-M_K)+1}
\end{equation}
we can estimate the physical distances $d$ [in pc] for our candidates if they were indeed carbon stars or M-giants. It shows that the minimum distances for these objects would be around $\sim$155 kpc for carbon stars and $\sim$135 kpc for late type M giants (Table~\ref{giant_magnitudes}). If the assumed extinction towards the objects (Table~\ref{candidates}) is not underestimated the derived distances lie way beyond the currently known extent of our Galaxy, and to our knowledge carbon stars have only been found at distances up to $\sim$130 kpc \citep[][and references therein]{mauron2007}. Only the minimum distance value for early M-type giants $\sim$60 kpc appears still reasonable, but in Figure~\ref{spectra1} we showed that at least CAHA Tau 2 is not an early M-type object. 

The density law of halo (giant) stars at those large Galacto-centric distances is largely unconstrained, so that it is difficult to estimate even roughly the number of such stars in our total data sample. Extrapolating the halo giant stellar count of \citet{Majewski03} assuming a smooth halo and a $r^{-2}$ surface density law, we expect much less than one object in our survey area. However, it is possible that the halo be highly structured at these distances, so that the counts would be dominated by (unknown) substructures. In particular, we note that Taurus lies on the Sagittarius stream orbital plane, on its Southern arc. However, no Sagittarius tails are known past $\approx 40$\,kpc, or predicted past 100\,kpc \citep{Law05}.

We think that Figure~\ref{cm_carbon} and Table~\ref{giant_magnitudes} provide additional arguments that most of our objects can be regarded as potential low-mass candidates. Especially the least luminous objects CAHA Tau 2 and 3 would be extremely distant if they were background objects. The only exception appears to be CAHA Tau 4. Already its photometric properties discussed in section~\ref{photometry} showed that this object might be of different nature than the other candidates. Figure~\ref{cm_carbon} further strengthens this hypothesis as CAHA Tau 4 could potentially be a (very) distant carbon star.

\subsection{Comparison to theoretical models}
Even though only spectra can eventually prove the young and low-mass nature of CAHA Tau 2 - 5 we provide in the following a first comparison of their photometric properties to theoretical models. The broadband SED of CAHA Tau 1 was already discussed above (Figure~\ref{caha1sed}) and that of CAHA Tau 6 (alias FU Tau B) was analyzed in \citet{luhman2009caha6}. For the other objects we derived estimates for effective temperatures, potential masses and surface gravities from the models of \citet{burrows2003,burrows2006} (for an assumed age of 3 Myr) and \citet{chabrier2000} (for an assumed age of 1 Myr). To derive these parameters we fitted the extinction corrected photometric values for the $I$, $R$ and $J$ band (Table~\ref{candidates}) to model grids covering a mass regime between 5 and 70 M$_{Jup}$. The $I$, $R$ and $J$ band were chosen for the following  reasons: (1) for these bands suitable model grids covering the mass regime were accessible for both ages; (2) focusing on the shorter wavelength regime of our observations reduces the potential contamination by circumstellar material altering the NIR fluxes; (3) the $J$-band photometry has a smaller intrinsic error than the optical wavelength bands and add an additional color to our fits\footnote{The derived best fit masses did not change for most objects when we used only the $I$ and $R$ band. Only for CAHA Tau 3 we found slightly lower masses in this case.}. For the extinction correction we used the individual values for $A_V$ given in Table~\ref{candidates} and applied the extinction law from \citet{mathis1990}. We furthermore assumed a distance of 140 pc (corresponding to a distance modulus of 5.73 mag) and adapted the photometric values of the theoretical models (originally given in absolute magnitudes) accordingly. 

Table~\ref{masses} summarizes the fit results \emph{assuming} that our objects are indeed young, low-mass objects. Changing the distance to 160 pc (performed as an additional test case) did only alter the results for one object, CAHA Tau 2, in the 1 Myr case. Here, a slightly higher mass was found compared to the value given in Table~\ref{masses}. Overall, this fitting exercise shows that, depending on the age, the first three objects would be among the least massive and coolest Taurus objects known today in case they were Taurus objects. In particular, CAHA Tau 2 could lie close to the deuterium burning limit and thus on the edge to planetary mass objects. However, as already mentioned in section~\ref{spect}, only spectra will yield the ultimate answer to spectral type and thus mass of our objects. 

\begin{figure}
\epsscale{1.}
\plotone{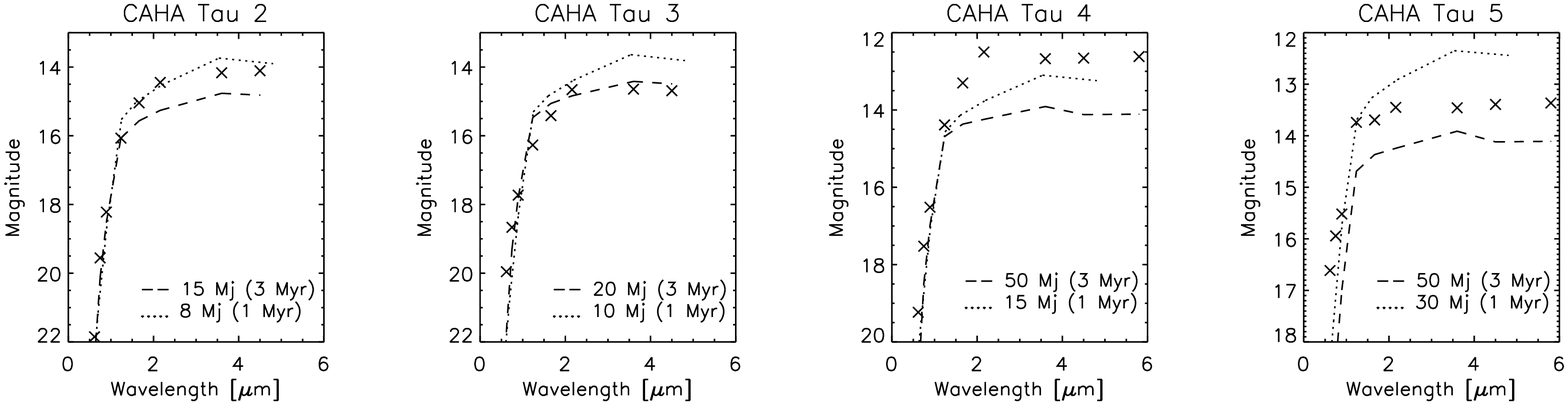}
\caption{Observed photometric values of the remaining candidates as listed in Table~\ref{candidates} compared to the best fit theoretical models shown in Table~\ref{masses}. The observed magnitudes were corrected for extinction using the extinction law of \citet{mathis1990} with $R_V$=3.1 and the estimated values for the objects' extinction given in Table~\ref{candidates}. \label{seds}}
\end{figure}

In Figure~\ref{seds} we compare all observed, extinction corrected magnitudes to the theoretical broadband SEDs of the best fit models given in Table~\ref{masses}. It shows that the magnitudes for CAHA Tau 2 and 4 at the shorter wavelength end are already in general agreement with the models while the results for CAHA 3 and 5 show notable deviations. Even if we imply that the objects are young, low-mass objects, as done in this Figure, then there are still two major sources of uncertainty: (1) the individual extinction can differ from the value derived from the extinction maps, and (2) the distance, and thus the required distance modulus, of the objects may vary. The plot for CAHA Tau 5, and partly also that for CAHA Tau 3, yield hints that the extinction might have been overestimated as the slope between the three optical SDSS bands is not as steep as required by any of the models. However, we emphasize again that only with the help of additional spectra the currently existing degeneracy between spectral type (and thus mass), $A_V$ and distance can be resolved and we refrain from further refinements of the exercise presented here. Also, the question of NIR/MIR excess emission due to a disk can only be answered once the spectral type and $A_V$ is confirmed for each object. For instance, while CAHA Tau 4 shows already with the current assumptions NIR excess emission, CAHA Tau 2 shows only excess emission compared to the 3 Myr models. 

\section{Conclusions}
Based on a deep NIR survey in Taurus we identified 6 new candidates for low-mass objects in this young star-forming region. One object, CAHA Tau 6 (now FU Tau B), was independently confirmed as young, low-mass BD of spectral type M9.25 by \citet{luhman2009caha6}. For a second object CAHA Tau 1, we have spectroscopic evidence that it is indeed a young, very low-mass objects and we assigned a spectral of L2$\pm$0.5, making it the object with the latest spectral type in Taurus. For the remaining 4 objects we thus far lack clear spectroscopic confirmation for their young, low-mass nature and we can only consider them as candidates at this point in time. However, the identification of CAHA Tau 1 and CAHA Tau 6 with spectral types L2 and M9.25, respectively, as well as the re-identification of 10 additional known low-mass Taurus members during the selection process, gives us confidence for the low-mass nature of the other objects. If they turn out to be indeed young Taurus sources then a comparison to theoretical models suggests that at least a second object (CAHA Tau 2) could be among the least massive known objects in this young star-forming region known today. In this case, its mass would be close to or below the deuterium burning limit of $\sim$13 M$_{Jup}$ and its effective temperature would lie below $\sim$2300 K making it also an L-type object.
The findings in detail:
\begin{itemize}
\item{CAHA Tau 1:}
The photometric properties, its proper motion and also the shape and spectroscopic features in the $J$-band spectrum make it very likely that CAHA Tau 1 is the first Taurus member with a spectral type as late as L2$\pm$0.5 and an effective temperature of $\sim$2100 K. For an age between 1 and 5 Myr, as derived from its position in a HR-diagram, this temperature corresponds to 5-15 M$_{Jup}$ and it might well be that CAHA Tau 1 is a free floating planetary mass object. Its broad band SED is consistent with model predictions for a 3 Myr-old object of 10 Jupiter masses with slight excess emission at 4.5\,$\mu$m possibly arising from a circum(sub-)stellar disk.  
\item{CAHA Tau 2:}
The optical and NIR photometric properties support the young, low-mass hypothesis for this object. Further evidence comes from its proper motion. A first $J$-band spectrum did not unambiguously prove the young, low-mass nature as the long wavelength end resembles that of evolved stars or carbon stars. However, a comparison of the photometric properties makes this rather unlikely. Assuming that CAHA Tau 2 is young, low-mass objects then theoretical models suggest that, depending on the age, it lies close to or below the deuterium burning limit making it another candidate for being among the least massive and coolest known Taurus members today. 
\item{CAHA Tau 3:} 
For CAHA Tau 3 we lack spectroscopic information but its photometric properties are in good agreement with those of young, very low-mass objects. However, currently the proper motion analysis does not support a Taurus membership. Additional data is needed to reduce the large error bars and get a more robust assessment of the proper motion. Assuming a young, low-mass nature then theoretical models suggest that CAHA Tau 3 lies somewhere at the border between M- and L-type objects with $\sim$15 M$\_{Jup}$. In this case, the currently applied value for the extinction is probably overestimated and needs to be revised which would again reduce the estimated mass.
\item{CAHA Tau 4 and 5:} 
The optical and NIR data for CAHA Tau 5 are in good agreement with young BD while CAHA Tau 4, the brightest object in our sample, lies a little bit off in the presented color-color and color-magnitude plots. Based on its photometric properties CAHA Tau 4 could also be a (very) distant, evolved object. A fit to theoretical models for young low-mass objects shows that if CAHA Tau 4 was a young BD then it would probably have significant NIR excess emission. Fitting models to the photometric values of CAHA Tau 5 reveals that the currently applied value for the extinction (and/or possibly also the applied distance modulus) need to be refined once spectroscopic data is available. Our PM analysis does not put conclusive constraints on the Taurus membership for both sources. 
\item{CAHA Tau 6:} This object was identified as FU Tau B, a young, M9.25 object in a BD binary system by \citet{luhman2009caha6} during the referee process of this paper. We publish here the first $J$, $H$ and $K_s$ band data for this source complementing the broadband SED from \citet{luhman2009caha6}.
\end{itemize}

So far we have not yet made an attempt to estimate the effect of our findings on the Taurus IMF. Once we know better which of the above mentioned candidates are indeed young, low-mass sources and how many additional candidates we find in the remaining and not yet here included fields of our deep NIR survey, it is certainly worth analyzing whether our results significantly influence the low-mass end of the IMF and help to fill-up the apparent deficiency of very low-mass objects compared to late K and early M type Taurus members.

\acknowledgments
This paper is dedicated to our colleague, friend and great pianist Dr. Frithjof Brauer who eventually lost his battle against his disease and passed away way too early. 

S. P. Q. kindly acknowledges support from the German
\emph{Friedrich-Ebert-Stiftung}. We are extremely grateful to the staff at ESO/Paranal and Calar Alto observatory for the great support during the observations. We kindly thank Stephan Birkmann, Micaela Stumpf, Steve Boudreault, Tigran Khanzadyan and Christoph Deist for their help during the observing runs on Calar Alto. We are indebted to Coryn Bailer-Jones and Viki Joergens for helpful discussions and to K. N. Allers for providing $J$-band reference spectra. We also thank an anonymous referee for valuable comments and suggestions that helped to improve the style and content of this work.
This research has made use of the SIMBAD database, operated at CDS, Strasbourg, France; 
and of the facilities of the Canadian Astronomy Data Centre operated by the National Research Council of Canada with the support of the Canadian Space Agency. This study was supported in part by NASA grant NNX07AG80G.


{\it Facilities:}  \facility{Calar Alto},\facility{VLT}, \facility{Spitzer}

\appendix
\section{Transformation equations}
\subsection*{2MASS - CIT}
(see, http://www.ipac.caltech.edu/2mass/releases/allsky/doc/sec6\_4b.html)
\begin{eqnarray}
\nonumber
(K_s)_{\rm 2MASS} = K_{\rm CIT} - (0.019\pm0.004)+(0.001\pm0.005)(J-K)_{\rm CIT}\\\nonumber
(J-H)_{\rm 2MASS}	 =	(1.087\pm0.013)(J-H)_{\rm CIT}	 - (0.047\pm0.007)\\\nonumber
(J-K_s)_{\rm 2MASS}	 =	(1.068\pm0.009)(J-K)_{\rm CIT}	 -	(0.020\pm0.007)\\\nonumber
(H-K_s)_{\rm 2MASS}	 =	(1.000\pm0.023)(H-K)_{\rm CIT}	+	(0.034\pm0.006)
\end{eqnarray}

\subsection*{2MASS - Bessel \& Brett}
(see, http://www.ipac.caltech.edu/2mass/releases/allsky/doc/sec6\_4b.html)
\begin{eqnarray}\nonumber
(K_s)_{\rm 2MASS}	 =	K_{\rm BB}	-	(0.039\pm0.007)	+	(0.001\pm0.005)(J-K)_{\rm BB}\\\nonumber
(J-H)_{\rm 2MASS}	 =	(0.990\pm0.012)(J-H)_{\rm BB}	 -	(0.049\pm0.007)\\\nonumber
(J-K_s)_{\rm 2MASS}	 =	(0.983\pm0.008)(J-K)_{\rm BB}	 -	(0.018\pm0.007)\\\nonumber
(H-K_s)_{\rm 2MASS}	 =	(0.971\pm0.022)(H-K)_{\rm BB}	+	(0.034\pm0.006)
\end{eqnarray}

\subsection*{SDSS (\emph{ugriz}) - Johnson-Cousins (\emph{UBVRI}) }
\citep[see,][]{jordi2006}
\begin{eqnarray}\nonumber
r - z = (1.584\pm0.008)(R-I) - (0.386\pm0.005)\\\nonumber
i-I = (0.247\pm0.003)(R-I) + (0.329\pm0.002)
\end{eqnarray}






\clearpage

\begin{deluxetable}{lcccc}
\tablecaption{Estimated masses of objects with an effective temperature of $\sim$2500 K for different ages based on theoretical models. Masses are given in Jupiter masses.
\label{bd_masses1}}           
\tablewidth{0pt}
\tablehead{
\colhead{Model}& \colhead{1 Myr}&\colhead{2 Myr} &\colhead{3 Myr} & \colhead{5 Myr}
}
\startdata
\citet{baraffe2003} &   $\sim$20 M$_{Jup}$ & - & - & $\sim$23 M$_{Jup}$\\
\citet{chabrier2000} &  $\sim$17 M$_{Jup}$ & - & - & $\sim$20 M$_{Jup}$\\
\citet{burrows2003,burrows2006} &  - &   $\sim$16 M$_{Jup}$ &  $\sim$17 M$_{Jup}$  & -\\
\enddata
\end{deluxetable}

\clearpage

\begin{deluxetable}{lcccc}
\centering
\tablecaption{Masses and effective temperatures for young, low-mass objects of different ages based on theoretical models. Masses are given in Jupiter masses.
\label{bd_masses2}}           
\tablewidth{0pt}
\tablehead{
\colhead{Model}& \colhead{1 Myr}&\colhead{2 Myr} &\colhead{3 Myr} & \colhead{5 Myr}
}
\startdata
\citet{baraffe2003} &   5 M$_{Jup}$ / 1900 K & - & - & 5 M$_{Jup}$ / 1410 K\\
				 &   10 M$_{Jup}$ / 2250 K & - & - & 10 M$_{Jup}$ / 1970 K\\
				 &    15 M$_{Jup}$ / 2400 K & - & - & 15 M$_{Jup}$ / 2260 K\\
\citet{chabrier2000} &   5 M$_{Jup}$ / 1900 K & - & - & 5 M$_{Jup}$ / 1410 K\\
				 &    10 M$_{Jup}$ / 2240 K & - & - & 10 M$_{Jup}$ / 1870 K\\
				 &    15 M$_{Jup}$ / 2450 K & - & - & 15 M$_{Jup}$ / 2250 K\\
\citet{burrows2003,burrows2006} & - &  5 M$_{Jup}$ / 1620 K & 5 M$_{Jup}$ / 1500 K & - \\
				 & - &  10 M$_{Jup}$ / 2130 K & 10 M$_{Jup}$ / 2040 K & - \\
				 & - &  15 M$_{Jup}$ / 2450 K & 15 M$_{Jup}$ / 2390 K & - \\		  
\enddata
\end{deluxetable}

\clearpage

\begin{deluxetable}{lcccccccc}
\centering
\tablecaption{Journal of the {\sc Omega2000} observations in Taurus. Column four shows the number of 1-minute exposures per field and filter that were finally co-added to create the final images. The seeing (column 5) was estimated during the data reduction in the respective NIR filters. The sixth column shows the mean transformation error for each field and filter resulting from the photometric calibration with 2MASS data.
\label{o2kobservations}} 

\tablewidth{0pt}
\tablehead{
\colhead{Field name}& \colhead{RA (J2000)}&\colhead{DEC (J2000)} & \colhead{\# of exposures} & \colhead{Seeing [$''$]} & \colhead{Calibr. err. [mag]}&\colhead{Obs. date} &\colhead{SDSS} \\
\colhead{}& \colhead{}&\colhead{} & \colhead{$J / H / K_s$} & \colhead{$J / H / K_s$} & \colhead{$J / H / K_s$} &\colhead{} &\colhead{coverage} 
}
\startdata
{ tc2} & { 04h10m53.64s} & { +$25^{\circ}09'28.3''$} & {  28 / 28 / 30}  & {  1.3 / 1.0 / 1.0}  & 0.04 / 0.05 / 0.05& { Jan. 05} & 50\% \\

{ tc5\_6\_7\_8} & { 04h18m00.87s}  & { +$28^{\circ}12'28.9''$} & {  30 / 28 / 30}  &  { 1.0 / 1.0 / 1.5}   & 0.07 / 0.05 / 0.05 & { Jan. 05} & 100\%\\
{  tc7\_11} & { 04h18m19.56s} & { +$28^{\circ}23'59.2''$} & { 30 / 30 / 30}  &  { 1.4 / 1.0 / 1.3}   & 0.09 / 0.07 / 0.07  & { Oct. 04} & 100\%\\
{ tc12a\_12b} & { 04h19m03.83s} & { +$27^{\circ}19'27.4''$} & { 30 / 30 / 21} & { 1.6 / 1.7 / 1.7}  & 0.05 / 0.03 / 0.06  & { Oct. 04} & ~92\% \\
{ tc13a\_13b\_14} & { 04h19m44.22s} & { +$27^{\circ}09'52.2''$} & { 30 / 27 / 29}  & { 1.2 / 1.0 / 1.2}   & 0.06 / 0.07 / 0.05 & { Jan. 05} & 100\%\\
{ tc15\_16a} & { 04h20m35.70s} & { +$27^{\circ}05'37.2''$} & {  30 / 30 / 30} &  { 1.4 / 1.3 / 1.3}   & 0.08 / 0.08 / 0.08 & { Dec. 05} & 100\% \\
{ tc16a\_16b\_17} & { 04h21m22.49s} & {+$26^{\circ}59'08.0''$} &{  30 / 30 / 30} & {  1.6 / 1.4 / 1.3}& 0.09 / 0.05 / 0.07  & { Dec. 05} & 100\%  \\
{ tc19} & { 04h23m36.53s} & { +$25^{\circ}05'55.3''$} &  { 30 / 30 / 30} &  { 1.3 / 1.3 / 1.2}& 0.06 / 0.04 / 0.05 & { Dec. 05} & 100\% \\
{ tc20\_21\_22} & { 04h23m55.75s} & { +$26^{\circ}36'38.4''$}  & { 30 / 30 / 30}  & { 1.8 / 1.7 / 1.2} & 0.05 / 0.05 / 0.06  & { Oct. 04} & 100\% \\
{ tc23\_24} & { 04h26m33.31s}  & { +$24^{\circ}38'47.4''$}  & {  30 / 30 / 29}  & { 1.4 / 1.5 / 1.3}  & 0.06 / 0.04 / 0.05 & { Oct. 04} & 100\% \\
{ tc25\_26a\_26b} & { 04h27m53.53s} & { +$26^{\circ}19'01.0''$} & {  30 / 30 / 30}  & { 1.2 / 1.6 / 1.2} & 0.04 / 0.06 / 0.05& { Dec. 05} & 100\% \\
{ tc28\_29} & { 04h29m40.00s} & { +$24^{\circ}29'34.1''$} &  { 30 / 30 / 28}  &  { 1.9 / 2.4 / 2.4} & 0.06 / 0.04 / 0.06 & { Oct. 04} & 100\%  \\
{ tc35\_36a\_36b} & { 04h35m44.37s} & { +$24^{\circ}09'11.8''$} & { 29 / 30 / 30}   & { 1.2 / 1.1 / 1.0} & 0.06 / 0.06 / 0.06& { Jan. 05} & 77\%   \\
{ tc37\_38\_39\_41}& { 04h39m28.62s} & { +$25^{\circ}47'29.6''$} & {  29 / 29 / 30}  &  { 1.0 / 1.0 / 1.1} & 0.09 / 0.07 / 0.05& { Jan. 05} & 100\%   \\
{ tc42a\_42b\_42c} & { 04h40m36.19s} & { +$25^{\circ}29'53.4''$} & {  29 / 30 / 30}  & { 0.9 / 1.0 / 0.9} & 0.07 / 0.05 / 0.05& { Jan. 05} & 100\%  \\
{ tc43\_44} & { 04h41m28.61s} & { +$25^{\circ}54'15.5''$} & {  30 / 29 / 30}  & { 1.0 / 1.1 / 1.0}   & 0.06 / 0.05 / 0.10& { Jan. 05} & 100\%  \\
{ tc50} & {  04h41m30.00s} & { +$25^{\circ}42'30.0''$} & {  30 / 30 / 30}  & { 1.2 / 1.2 / 1.1} & 0.07 / 0.06 / 0.06 &{  Dec. 05/Jan. 06} & 100\%  \\
\hline \\
{ tc1a\_1b} &{  04h04m46.45s} & { +$26^{\circ}19'27.7''$} &{  30 / 43 / 30}  & { 1.7 / 1.4 / 1.3} & 0.11 / 0.05 / 0.06 & { Dec. 05} & -  \\
{ tc3\_4} &{ 04h13m52.10s} & { +$28^{\circ}15'45.8''$}  & { 30 / 30 / 30}  & { 1.0 / 1.1 / 1.0} & 0.05 / 0.04 / 0.03& { Jan. 05} & -  \\
{ tc9\_10a\_10b} & { 04h18m25.60s} & { +$27^{\circ}30'01.3''$} & { 30 / 30 / -}  &  { 2.3 / 1.8 / -}   & 0.08 / 0.04 / -& { Dec. 05} & -\\
{ tc18a\_18b} & { 04h21m58.98s} & { +$15^{\circ}30'02.4''$} & { 30 / 29 / 28} & { 1.1 / 1.2 / 1.3}& 0.04 / 0.04 / 0.06 & { Jan./Feb. 05} &-  \\
{ tc27} & { 04h28m39.26s }& { +$26^{\circ}51'39.2''$}  & {  30 / 30 / 30}  & { 1.5 / 1.3 / 1.3}  & 0.07 / 0.07 / 0.06& { Dec. 05} & - \\
{ tc30} & { 04h31m41.90s}  & { +$18^{\circ}09'50.7''$}  & {  30 / 30 / 30}    &  { 2.6 / 1.9 / 2.4} & 0.07 / 0.06 / 0.05& { Oct. 04} & -  \\
{ tc31\_32\_33a\_33b} & { 04h32m19.18s }& { +$24^{\circ}28'00.6''$} & { 30 / 30 / -}  &  { 1.2 / 1.3 / -} & 0.06 / 0.06 / - & { Dec. 05} & - \\
{ tc34} & {  04h33m24.28s } & { +$22^{\circ}42'15.9''$} & {  28 / 29 / 27} & { 1.2 / 1.2 / 1.2}   & 0.06 / 0.06 / 0.07& { Jan./Feb. 05} & -\\
\enddata
\end{deluxetable}

\clearpage

\begin{deluxetable}{lccccccc}
\centering
\tablecaption{Properties of selected but already known low-mass objects in Taurus. For each object magnitudes from our observations as well as from the 2MASS catalogue (second line of magnitudes) are given for comparison. The errors in our NIR magnitudes reflect uncertainties in the fitting of the individual PSFs as well as systematic errors resulting from the photometric calibration with the 2MASS reference sources. References: (a) \citet{luhman2006spitzer}, (b) \citet{luhman2006distribution}, (c) \citet{guieu2006}, (d) \citet{white2003}, (e) \citet{luhman1998},
(f) \citet{mohanty2005}, (g) \citet{harris1988}, (h) \citet{itoh1999}, (i) \citet{itoh2002}, (j) \citet{briceno2002}, (k) 2MASS Point Source Catalogue \citep{cutri2003}, (l) \citet{luhman2009caha6}
\label{known_objects}}           
\tablewidth{0pt}
\tablehead{
\colhead{Object}& \colhead{RA (J2000)}&\colhead{DEC (J2000)} & \colhead{$J$} & \colhead{$H$} &\colhead{$K_s$} &\colhead{Spec. Type}&  \colhead{Ref.}
}
\startdata
V410 X-ray 3  &   04h18m08.0s  & $+28^{\circ}26'03.9''$  & 11.60$\pm$0.10 & 10.91$\pm$0.09 & 10.52$\pm$0.09 & M6-M6.5   &  (d),(e),(f)\\
			&			&					& 11.54$\pm$0.02 & 10.82$\pm$0.02 & 10.45$\pm$0.02 & & (k)\\
IRAS 04154+2823  & 04h18m32.1s  &  $+28^{\circ}31'15.5''$ &  14.38$\pm$0.10 & 12.27$\pm$0.10 &  9.59$\pm$0.08 &-    & (g) \\
				&				&				&15.19$\pm$0.05 & 12.37$\pm$0.02 & 10.27$\pm$0.02	&    & (k)\\
 2MASS J04242090+263051 &  04h24m20.9s &  $+26^{\circ}30'51.0''$ & 13.40$\pm$0.06 & 12.82$\pm$0.06 & 12.48$\pm$0.07 & M6.5  &  (a) \\
 					&				&					& 13.49$\pm$0.02 & 12.81$\pm$0.02 & 12.43$\pm$0.02 &  & (k)\\	
2MASS J04263055+2443558 &  04h26m30.6s & $+24^{\circ}43'55.8''$ &  14.74$\pm$0.07 & 13.95$\pm$0.05 & 13.40$\pm$0.06 & M8.75  &  (b)\\
					&				&					& 14.67$\pm$0.03 & 13.94$\pm$0.04 & 13.40$\pm$0.03 & & (k)\\
 CFHT-BD-Tau 20 &  04h29m59.5s & $+24^{\circ}33'07.7''$ & 11.77$\pm$0.07 & 10.89$\pm$0.05 & 10.14$\pm$0.07 & M5.5  &  (c)\\
 				&			& 				& 11.68$\pm$0.02 &10.54$\pm$0.03 & 9.81$\pm$0.02 && (k)\\
 ITG33           &      04h41m08.3s & $+25^{\circ}56'07.3''$ &  14.01$\pm$0.07 & 12.51$\pm$0.08 & 11.62$\pm$0.13 & K7-M3   &   (h),(i)\\
 			& 				& 				& 13.74$\pm$0.03 & 12.15$\pm$0.02 & 11.09$\pm$0.02	& & (k)\\
 CFHT-BD-Tau 19    &  04h21m08.0s & $+27^{\circ}02'20.1''$ & 12.00$\pm$0.10 & 10.33$\pm$0.09 & 9.19$\pm$0.12 & M5.25  & (c)\\
 				&			& 					& 13.86$\pm$0.02 & 12.06$\pm$0.03 & 10.54$\pm$0.02 & & (k)\\
 2MASS J04215450+2652315  &   04h21m54.5s &  $+26^{\circ}52'32.0''$ &  15.53$\pm$0.10 & 14.46$\pm$0.06 & 13.81$\pm$0.09 & M8.5 &  (b) \\          
 						&				& 				& 15.54$\pm$0.04 &	14.50$\pm$0.04 & 13.90$\pm$0.04 & & (k)\\	
FU Tau  A &   04h23m35.4s &  $+25^{\circ}03'02.5''$ &  10.77$\pm$0.07 & 9.98$\pm$0.06 & 9.45$\pm$0.07 & M7.25  & (g)/(l) \\
		&			&					& 10.78$\pm$0.03 & 9.95$\pm$0.03 & 9.32$\pm$0.02 & &(k)\\
 KPNO-Tau 4	      &   04h27m28.0s & $+26^{\circ}12'04.7''$ & 15.06$\pm$0.05 & 13.97$\pm$0.07 & 13.29$\pm$0.06 & M9.5   & (j)\\
 				&			&					& 15.00$\pm$0.04 & 14.03$\pm$0.04 & 13.28$\pm$0.03 & & (k)\\
\enddata


\end{deluxetable}

\clearpage



\begin{deluxetable}{cccccccc}
\tablecaption{Coordinates and photometry of the newly identified candidates for young (very) low-mass objects in Taurus as well as rough estimates for their optical extinction. 
Optical magnitudes are given in the SDSS photometric system ($r$, $i$, $z$) and Johnson-Cousin system ($R$,$I$), the NIR magnitudes ($J$, $H$, $K_s$) are in the 2MASS system, and the 
longest wavelengths given here are Spitzer/IRAC magnitudes. The errors in the $R$ and $I$ band are derived from the transformation equation (see appendix) and the individual errors in the SDSS magnitudes. The NIR errors account for transformation errors between the {\sc Omega2000} filter system and the 2MASS photometric system as well as for individual fitting errors during the PSF photometry. The errors in the IRAC bands reflect the uncertainties in the aperture photometry.  
\label{candidates}}           
\tablewidth{0pt}
\tablehead{
\colhead{Object}&\colhead{RA (J2000)}&\colhead{DEC (J2000)}&\colhead{$r$}&\colhead{$i$}& \colhead{$z$}&\colhead{$R$}&\colhead{$I$}
}
\startdata
CAHA Tau 1& 04h36m09.0s  &  $+24^{\circ}08'21.2''$ & 23.42$\pm$0.41 &  21.09$\pm$0.08 & 19.34$\pm$0.07  & 22.89$\pm$0.38 & 20.00$\pm$0.35\\

CAHA Tau 2 &    04h39m47.2s &    $+25^{\circ}53'32.2''$  &    22.65$\pm$0.22   &  20.17$\pm$0.04 & 18.66$\pm$0.04 & 21.92$\pm$0.22 & 19.16$\pm$0.17 \\

CAHA Tau 3  &   04h40m39.3s  &   $+25^{\circ}23'02.0''$ &  23.60$\pm$0.48 & 21.47$\pm$0.11 & 19.73$\pm$0.12 & 23.16$\pm$0.57 & 20.47$\pm$0.47\\

CAHA Tau 4  &    04h40m57.4s &    $+25^{\circ}50'07.9''$ &  23.41$\pm$0.41 &  20.74$\pm$0.07& 18.81$\pm$0.05 & 22.78$\pm$0.41 & 19.63$\pm$0.31\\  

CAHA Tau 5   &    04h41m31.6s &    $+25^{\circ}49'32.2''$   & 20.17$\pm$0.03 &  18.68$\pm$0.01 & 17.47$\pm$0.02 & 19.82$\pm$0.05 & 17.87$\pm$0.04\\  


CAHA Tau 6\tablenotemark{a}  &    04h23m35.8s &    $+25^{\circ}02'59.4''$  &   22.25$\pm$0.14   &  20.19$\pm$0.05 & 17.80$\pm$0.05 & 22.16$\pm$0.22 & 19.11$\pm$0.20\\ 
\\
\hline\hline

 & $J$ & $H$ & $K_s$ & 3.6 $\mu$\,m & 4.5 $\mu$\,m & 5.8 $\mu$\,m & $A_V$\tablenotemark{d} \\
\\
\hline

CAHA Tau 1 &  16.92$\pm$0.07 & 15.77$\pm$0.09 & 15.11$\pm$0.09 & 14.62$\pm$0.08 & 14.43$\pm$0.09  & - & $\sim 2.3$  \\

CAHA Tau 2  & 16.33$\pm$0.12 & 15.20$\pm$0.10 & 14.55$\pm$0.08 &  14.21$\pm$0.07\tablenotemark{b} & 14.14$\pm$0.08\tablenotemark{b} & - & $\sim 0.9$\\

CAHA Tau 3   & 17.46$\pm$0.10  &  16.13$\pm$0.08 & 15.12$\pm$0.10 & 14.84$\pm$0.08 & 14.83$\pm$0.11 & - & $\sim 4.1$\\

CAHA Tau 4  &  15.75$\pm$0.07& 14.12$\pm$0.08 & 13.04$\pm$0.17 & 12.90$\pm$0.06 & 12.82$\pm$0.06 & 12.73$\pm$0.16 & $\sim 4.7$\\  

CAHA Tau 5     &  14.90$\pm$0.14 & 14.39$\pm$0.08 & 13.90$\pm$0.08 & 13.65$\pm$0.06 & 13.53$\pm$0.07 & 13.46$\pm$0.26 &$\sim 4.0$\\  


CAHA Tau 6\tablenotemark{a} &  15.10$\pm$0.08 &  14.12$\pm$0.05 &  13.43$\pm$0.07 & 12.54$\pm$0.1\tablenotemark{c}    &  11.90$\pm$0.1\tablenotemark{c}   & 11.46$\pm$0.1\tablenotemark{c} &  $\sim 4.9$/$<1.0$\tablenotemark{c}\\
\enddata
\tablenotetext{a}{Identified as FU Tau B by \citet{luhman2009caha6} during the revision process of this paper.}
\tablenotetext{b}{Object lies in so-called ''pull-down'' column from a bright star in the IRAC images. IRAC fluxes are probably underestimated.}
\tablenotetext{c}{Figures taken from \citet{luhman2009caha6}.}
\tablenotetext{d}{Values are estimates based on the Taurus extinction map by \citet{dobashi2005} which is created from DSS1 optical data with a resolution of $\sim 6'$.}
\end{deluxetable}

\clearpage





\begin{deluxetable}{lccc}
\tablecaption{Proper motions and time baseline of our candidates.
\label{PMcand}}           
\tablewidth{0pt}      
\tablehead{
\colhead{}& \colhead{$\mu_\alpha$ [mas/yr] } &\colhead{$\mu_\delta$ [mas/yr]} & \colhead{time baseline  [d] }}
\startdata
  CAHA Tau 1 &  $+12\pm 20$ & $-26\pm 15$ & 1373 \\ 
 CAHA Tau 2 & $+10\pm 15$  & $-28\pm 10$ & 1566    \\
 CAHA Tau 3 &   $-42\pm 40$ & $+7\pm 20$  & 1355  \\
 CAHA Tau 4 & $+29\pm 20$ &  $-17\pm 30$&  2270 \\
 CAHA Tau 5 & $-8\pm 16$ & $-10\pm 16$ & 1514 \\
\enddata
\end{deluxetable}

\clearpage

\begin{deluxetable}{cccccc}
\tablecaption{Properties of CAHA Tau 1 based on its J-band spectrum.
\label{caha1_table}}           
\tablewidth{0pt}      
\tablehead{
\colhead{Spectral type}& \colhead{T$_{eff}$ [K]\tablenotemark{a}} &\colhead{BC$_K$\tablenotemark{a}} & \colhead{log L/L$_\sun$\tablenotemark{b}} & \colhead{Age [Myr]} & \colhead{Mass [$M_{Jup}]$\tablenotemark{c}}\\
}
\startdata
L2$\pm$0.5 & 2080$\pm$140   &  3.3$\pm$0.2 & 3.1$\pm$0.1 & 1 / 3 / 5 & 7$^{+3}_{-2}$ / 10$^{+2}_{-2}$ / 12$^{+3}_{-2}$\\
\enddata
\tablenotetext{a}{Based on the relations between, SpT, T$_{eff}$ and bolometric correction in the K band (BC$_K$) found by \citet{golimowski2004}.}
\tablenotetext{b}{Derived using: $L = 3.02\cdot10^{28}\times10^{-0.4\cdot M_{bol}}$ [W] \citep{lang1980}.}
\tablenotetext{c}{Most likely mass range for an age of 1, 3 and 5 Myr based on models of \citet{chabrier2000} and \citet{burrows2003,burrows2006}.}
\end{deluxetable}

\clearpage

\begin{deluxetable}{lcccc}
\tablecaption{Typical absolute K-band magnitudes of carbon stars and M-type giants and corresponding minimum and maximum distance moduli and distances if our remaining two least luminous candidates (CAHA Tau 2 and 3) were such objects. CAHA Tau 2 serves as basis for the minimum estimate ($m_K=14.45$), CAHA Tau 3 for the maximum ($m_K=14.67$)\tablenotemark{a}. References: (a) \citet{liebert2000,claussen1987}, (b) SIMBAD Astronomical Database (2MASS catalogue, Hipparcos catalogue): HD 196610, HD 18191, HD 207076, HD 108849; (c) SIMBAD Astronomical Database (2MASS catalogue, Hipparcos catalogue): HD 213893, HD 204724, HD 120052, HD 219734
\label{giant_magnitudes}}           
\tablewidth{0pt}      
\tablehead{
\colhead{Object type}& \colhead{M$_K$ [mag]} &\colhead{Reference} & \colhead{m$_K$-M$_K$ [mag]\tablenotemark{a}} & \colhead{Distance [kpc]}\\
\colhead{}& \colhead{min / max}&\colhead{}&\colhead{min / max}&\colhead{min / max}
}
\startdata
Carbon stars   &  -6.5 / -8.5   &  (a) & 20.95 / 23.17 &  155 / 431 \\
M6 III - M8 III &  -6.2 / -7.4  & (b)  & 20.65 / 22.07 & 135 / 259 \\
M0 III - M2.5 III & -4.2 / -6.0 &  (c) & 18.65 / 20.67 &  54  / 136 \\
\enddata
\tablenotetext{a}{The apparent $K$-band magnitudes of CAHA Tau 2 and 3 were corrected for extinction using the optical extinction values given in Table~\ref{candidates} and assuming $A_K/A_V\approx 0.11$ \citep{mathis1990}.}
\end{deluxetable}

\clearpage

\begin{deluxetable}{cccccc}
\tablecaption{Best fit results for mass, effective temperature and surface gravity of our candidates for 1 and 3 Myr. The figures for the extinction are the same as in Table~\ref{candidates}. The left figure in each column reflects the best fit to the 1 Myr models by \citet{chabrier2000}, the right figure in each column reflects the best fit to the 3 Myr models by \citet{burrows2003, burrows2006}. The overall better fit to the models is indicated by the asterisk in the 'Age' column.
\label{masses}}           
\tablewidth{0pt}      
\tablehead{
\colhead{Object} & \colhead{$A_V$ [mag]} & \colhead{Age [Myr]} & \colhead{Mass [M$_{Jup}$]}& \colhead{T$_{eff}$ [K]}& \colhead{log g}}
\startdata
CAHA Tau 2 &  0.9 & 1\tablenotemark{*} / 3 & 8 / 15 &  2159 / 2399 &  3.5 / 4.0  \\
CAHA Tau 3 & 4.1 & 1 / 3\tablenotemark{*} & 10 / 20 & 2240 /  2637 & 3.6 / 3.9 \\
CAHA Tau 4 & 4.7 & 1\tablenotemark{*} / 3 & 15 / 50 & 2337 / 2902 & 3.6 / 3.6 \\
CAHA Tau 5 & 4.0 & 1\tablenotemark{*} / 3 &  30 / 50 & 2703 / 2902 & 3.7 / 3.6 \\
\enddata
\tablenotetext{*}{Best fit model.}
\end{deluxetable}

\clearpage



\clearpage

\clearpage

\clearpage

\clearpage

\clearpage

\clearpage

\clearpage

\clearpage

\clearpage

\clearpage

\clearpage

\clearpage

\end{document}